\documentclass[11pt,a4paper]{article}
\usepackage{a4wide}
\usepackage{lineno,hyperref}
\usepackage{booktabs}
\usepackage{amsmath}
\usepackage{amssymb}
\usepackage{bbold}
\usepackage{wasysym}
\usepackage[flushleft]{threeparttable}
\usepackage{lscape}
\usepackage{rotating}
\usepackage{upgreek}
\usepackage[english]{babel}
\usepackage{multirow}
\usepackage{wrapfig}
\usepackage[section]{placeins}
\usepackage[backend=biber,url,doi=false,giveninits=true,hyperref,style=numeric,citestyle=numeric-comp,sorting=none]{biblatex}

\usepackage{bm}
\usepackage{booktabs}
\usepackage{tikz}
\usetikzlibrary{decorations.markings}
\usepackage{graphicx}
\usepackage[capitalise]{cleveref}
\usepackage{siunitx}
\usepackage[colorinlistoftodos,prependcaption,textsize=tiny,linecolor=red,backgroundcolor=red!25,bordercolor=red]{todonotes}

\newcommand{\norm}[1]{\lVert#1\rVert}

\def\rms{{\rm s}}

\newcommand{\df}[1]{\textit{#1}}

\def\sl{\rms_\lambda}

\def\proof{\noindent{\sl Proof:}\kern0.6em}

\def\dual{\mathstrut^*\kern-0.1em}

\def\lvec#1{\setbox0=\hbox{$#1$}
    \setbox1=\hbox{$\scriptstyle\leftarrow$}
    #1\kern-\wd0\smash{
    \raise\ht0\hbox{$\raise1pt\hbox{$\scriptstyle\leftarrow$}$}}
    \kern-\wd1\kern\wd0}
\def\rvec#1{\setbox0=\hbox{$#1$}
    \setbox1=\hbox{$\scriptstyle\rightarrow$}
    #1\kern-\wd0\smash{
    \raise\ht0\hbox{$\raise1pt\hbox{$\scriptstyle\rightarrow$}$}}
    \kern-\wd1\kern\wd0}
\def\slash#1{\setbox0=\hbox{$#1$}\setbox1=\hbox{$\kern1pt/$}
    #1\kern-\wd0\kern1pt/\kern-\wd1\kern\wd0}

\def\nabstar#1{{\nabla\kern0.5pt\smash{\raise 4.5pt\hbox{$\ast$}}
               \kern-5.5pt_{#1}}}

\def\nabbarstar#1{{\overleftarrow{\nabla}\kern0.5pt\smash{\raise 4.5pt\hbox{$\ast$}}
               \kern-5.5pt_{#1}}}

\def\nabdbarstar#1{{\overleftrightarrow{\nabla}\kern0.5pt\smash{\raise 4.5pt\hbox{$\ast$}}
               \kern-5.5pt_{#1}}}

\def\drvstar#1{{\partial\kern0.5pt\smash{\raise 4.5pt\hbox{$\ast$}}
               \kern-6.0pt_{#1}}}

\def\ldrvstar#1{{\lvec{\,\partial}\kern-0.5pt\smash{\raise 4.5pt\hbox{$\ast$}}
               \kern-5.0pt_{#1}}}

\def\fm{{\rm fm}}
\def\MSbar{\overline{\rm MS\kern-0.5pt}\kern0.5pt}
\def\Nf{{N_{\rm f}}}
\def\Nb{V_1}
\def\Nc{{N_{\rm c}}}
\def\Ns{{N_{\rm s}}}
\def\Nsrc{{N_{\eta}}}
\def\Nl{{N_{\ell}}}

\def\zetabar{\bar{\zeta}}
\def\zetaprime{\zeta\kern1pt'}
\def\zetabarprime{\zetabar\kern1pt'}

\def\diracstar#1#2{
    \setbox0=\hbox{$\gamma$}\setbox1=\hbox{$\gamma_{#1}$}
    \gamma_{#1}\kern-\wd1\kern\wd0
    \smash{\raise4.5pt\hbox{$\scriptstyle#2$}}}

\def\tr{{\rm tr}}

\def\Ds{D_{\rm s}}
\def\DsdagDs{\Ds{\Ds}^{\kern-1pt\dagger}}

\def\avg#1{{\kern1.0pt\overline{\kern-1.0pt#1\kern-1.0pt}\kern1.0pt}}

\newcommand{\be}{\begin{equation}}
\newcommand{\ee}{\end{equation}}
\newcommand{\bea}{\begin{eqnarray}}
\newcommand{\eea}{\end{eqnarray}}

\newcommand{\msbar}{{\rm \overline{MS\kern-0.05em}\kern0.05em}}

\newcommand{\ba}{\begin{eqnarray}}
\newcommand{\ea}{\end{eqnarray}}

\newcommand\vslattice{\mathcal{V}}  

\newcommand\ispacetime{\Lambda}     

\newcommand{\cost}[1]{\text{cost}(#1)}
\newcommand{\iter}[1]{\text{iter}(#1)}
\newcommand{\mem}[1]{\text{mem}(#1)}

\DeclareSymbolFont{extraitalic}      {U}{zavm}{m}{it}
\DeclareMathSymbol{\Qoppa}{\mathord}{extraitalic}{161}
\newcommand{\ratio}{\Phi} 

\newcommand{\projector}{P}    
\newcommand{\gDop}{Q}         

\newcommand{\prop}{S}                   
\newcommand{\propi}[1]{S_{#1}}          

\newcommand{\innerprod}[2]{\left(#1, #2\right)} 

\usepackage{xparse} 
\ExplSyntaxOn
\NewDocumentCommand{\coarse}{m}{ 
    \int_compare:nTF { \tl_count:n { #1 } > 1 }
        { \widehat{#1} }
        { \hat{#1} }
}
\ExplSyntaxOff

\crefname{table}{Tab.}{Tabs.}
\Crefname{table}{Table}{Tables}
\crefname{section}{Sec.}{Secs.}
\Crefname{section}{Section}{Sections}
\crefname{appendix}{App.}{Apps.}
\Crefname{appendix}{Appendix}{Appendices}

\begin{document}
\begin{titlepage}
\begin{flushright}
\end{flushright}

\begin{center}
    
$\;\;\;$
\vspace{1.5cm}

{\Large\bf Multigrid low-mode averaging}

\end{center}
\vskip 0.75 cm
\begin{center}
{\large 
Roman Gruber, Tim Harris, Marina Krstić Marinković
\vskip 1.0cm
Institute for Theoretical Physics, ETH Z\"urich,\\
Wolfgang-Pauli-Str.~27, 8093 Z\"urich, Switzerland
}
\vskip 0.5cm
{\bf Abstract}
\vskip 1.5ex
\end{center}

\noindent

We develop a generalization of low-mode averaging in which the number of low
quark modes of the Dirac operator required for a constant variance reduction
can be kept independent of the volume by exploiting their local
coherence.
Typically in lattice QCD simulations, the benefit of translation averaging
quark propagators over the space-time volume is spoiled by large fluctuations
introduced by the approximations needed to estimate the average.
For quark-line connected diagrams at large separations, most of this
additional variance can be efficiently suppressed by the introduction of
hierarchical subspaces, thanks to the reduced size of the coarse grid operators
that act within the subspaces.
In this work, we investigate the contributions to the variance of
the isovector vector current correlator with $\Nf=2$ non-perturbatively
$\mathrm O(a)$-improved Wilson fermions on lattices approximately of size
$L=2,3$ and $4\,\fm$.
The numerical results obtained confirm that the variance decreases as the volume is increased 
when a multigrid decomposition is used with a fixed number of low modes.
While the proposed decomposition can be applied to any quark propagator, it
is expected to be especially effective for quark-line connected diagrams at
large separations, for example, the isovector contribution to the hadronic
vacuum polarization or baryonic correlators.

\vfill

\end{titlepage}


\section{Introduction}
\label{sec:intro}

Numerical simulations of lattice quantum chromodynamics (QCD) are an essential
tool to study the low-energy dynamics of the strong interactions and have
entered the precision era by providing input on the fundamental parameters of
the Standard Model, the hadronic spectrum and matrix
elements~\cite{FlavourLatticeAveragingGroupFLAG:2021npn,BMW:2014pzb,Aoyama:2020ynm}.
In order to control systematic uncertainties, the regularizations provided by
the finite lattice spacing $a$ and finite lattice size $L$ have to be removed
carefully.
Although a solid theoretical description of these limits is provided by
various effective field theory frameworks, in practice it is necessary to span
a wide range of lattice spacings and volumes to ensure the relevant scaling
regime is reached.
Crucial to this program is then the precise computation of correlation
functions close to the continuum and in large volume.

As is well known, when the continuum limit is approached, several issues
obstruct the precise determination of hadronic observables.
In addition to the critical slowing down of simulation algorithms like the
hybrid Monte Carlo algorithm~\cite{Schaefer:2010hu}, the statistical
uncertainty of many quantities increases with fixed
statistics, see e.g.~\cite{Giusti:2019kff,Altenkort:2021jbk}.
In this work, we are concerned with the scaling of the
variance, the squared statistical uncertainty, of hadronic correlation
functions as the volume is increased, at fixed lattice spacing.
For suitably localized observables averaged over the space-time lattice, the
variance decreases inversely proportional to the lattice volume $V=L_0\times
L^3$ in large enough volumes~\cite{Luscher:2017cjh}.
Therefore, in addition to reducing the systematic uncertainty, large volumes
can also be employed to reduce the statistical uncertainty, which is utilized in the master-field scenario~\cite{Francis:2019muy}.

In intermediate volumes, or regions of a master-field lattice, the evaluation
of the lattice average of fermionic observables can still be costly.
In order to avoid a cost growing with $V^2$, the translation averages are
often themselves estimated stochastically by introducing auxiliary random
fields~\cite{hutch_1990, Michael:1998sg} or sampled in other ways.
Such estimators tend to introduce large additional sources of variance,
which need to be suppressed so that translation averaging can be fully
exploited.
Efficient estimators for the quark propagator at small distances, relevant for the traces of
single-quark propagators that appear in correlators involving singlet fermion
bilinears, can be constructed by changing the quark mass
parameter~\cite{ETM:2008zte,Dinter:2012tt,Giusti:2019kff}, for example.

As the separation in the quark propagator is increased, the regime relevant for
quark-line connected diagrams at large distances, both the signal and the
variance are eventually dominated by the low quark modes.
This allows the creation of efficient estimators by computing a relatively small
number of low modes of the Dirac operator
exactly~\cite{Neff_2001,Giusti_2004,DeGrand_2004}, which we refer to as low-mode
averaging (LMA).
However, as the mode density grows proportional to the volume, the number
required for a constant variance reduction also increases~\cite{banks1980} and
may become prohibitively large even in intermediate volumes.
For high-precision determinations of the isovector contribution to the hadronic
vacuum polarization (HVP) of the muon $g-2$, for example, this may easily reach $\mathrm
O(1000)$ low modes per configuration on state-of-the-art
lattices~\cite{RBC_2024, bmw_2017, Djukanovic:2024cmq, bmw_2024, milc_gm2,
ExtendedTwistedMass:2022jpw, FermilabLattice:2024yho}.
If the number of modes is not sufficient then the fluctuations of the
remaining terms are still large, which is often observed in
practice~\cite{Djukanovic:2024cmq}.
Ultimately low-mode averaging alone does not solve the problem of the cost of
implementing translation averaging in the regime of large volumes.

A solution can be found by exploiting the property of the local coherence of
the low-quark modes~\cite{Luescher2007}, also known as the weak approximation
property~\cite{babich2010}.
That is, in local regions of space-time the low-mode space is to a good
approximation spanned by a small number of local fields.
This allows one to build a good approximation to the low-mode subspace
starting with a very small number of global modes and locally project them,
thereby spanning a much larger space than that defined by the
original basis.
This property is the foundation of efficient deflated
solvers~\cite{Luescher2007} and multigrid preconditioning~\cite{babich2010,Frommer:2013,Alexandrou:2016izb,Brower_2018,Brower:2020xmc,Boyle:2021wcf, Espinoza-Valverde:2022pci,Ayyar:2022krp, Boyle:2024pio}
for the Dirac equation, while also being used to accelerate the generation of
low-modes and reduce their storage requirements~\cite{Clark_2018}.

In this work, we follow an approach where we use block-decomposed low modes
directly in the decomposition of the quark propagator in the observable, which
allows us to span a much larger subspace than using the global low modes alone.
The dimension of coarse operator associated with the subspace should still be smaller
than the full Dirac operator and therefore the corresponding variance can be
efficiently suppressed.
Similar decompositions have been put forward and studied
Refs.~\cite{Bali:2015qya,ROMERO2020109356,Frommer:2021tqd} for single-propagator
traces relevant to quark-line disconnected diagrams.
Inspection of the corresponding variances suggests that low-mode averaging may find a more promising application to quark-line connected correlation functions at large separations~\cite{Gruber:2024czx,Giusti:2019kff}.
Here, we apply the decomposition to the quark-line connected isovector vector
correlator at large separations of the currents and show that a much reduced
cost for the translation average can be achieved compared to a simple stochastic
estimator.
A closer investigation of the proposed algorithm shows mild scaling with lattice
volume which vastly improves the computational and storage overhead compared to
low-mode averaging.
We argue that the favourable scaling is due to the subspace dimension being
proportional to volume.

The rest of this article is arranged as follows. We introduce a two-level decomposition of the quark propagator in \cref{sec:deflation} and
review the simplest conceivable scheme of low-mode averaging for the isovector
vector correlator as an example.
In \cref{sec:loccoh}, we review the property of local coherence and discuss the
relevant scales of the block decomposition pertinent to lattice QCD based on
some preliminary numerical tests.
After this in \cref{sec:mglma} we introduce the block decomposed version and its extension to a
hierarchical method which we call multigrid low-mode averaging.
In \cref{sec:results} we investigate the proposed method applied to the
isovector vector correlator in the time-momentum representation at different
separations and illustrate the volume scaling of the variance, using $\Nf=2$
non-perturbatively $\mathrm O(a)$-improved Wilson fermions.
Finally, in \cref{sec:conclusion}, we provide some discussion in the conclusions of future directions in
which the algorithm could be developed and applied.

\section{Variance reduction using deflation}
\label{sec:deflation}

In this section we review the main ingredients of a simple implementation
of low-mode averaging, which serves to introduce some notation and the main
challenges encountered.
First we introduce a general decomposition of the quark propagator
$S(x,y)$, the Green's function for the lattice Dirac operator $D$,
that forms the basis of low-mode averaging and its variants.
The precise form of the lattice discretization is not yet important and will
be specified later.
The decomposition is based on defining an arbitrary subspace of the space of
quark fields $\vslattice_0$.
This subspace can be taken to be spanned by the set
$\phi_0(x),\ldots,\phi_{\Nc-1}(x)$ which are assumed at this stage only to be
orthonormal.
The choice of the basis fields and its implications will be discussed in
\cref{sec:lma} and \cref{sec:loccoh}.

For convenience, let us define a linear operator $R: \vslattice_0\rightarrow \vslattice_1$ called the
restrictor from the space of complex quark fields of dimension $12V/a^4$ to the
space of complex coarse fields of dimension $\Nc$.
The space of coarse fields is then clearly isomorphic to the subspace of
$\vslattice_0$ spanned by the $\Nc$ basis fields.
It is defined by its action on a quark field $\psi_0 \in \vslattice_0$ by
\begin{align}
    (R\psi_0)_c = \sum_x\phi^\dagger_c(x)\psi_0(x),\qquad c=0,\ldots,\Nc-1.
\end{align}
In addition we define the prolongator $T: \vslattice_1\rightarrow \vslattice_0$ from the space of coarse
fields to quark fields defined by its action on a coarse field $\psi_1 \in \vslattice_1$ as
\begin{align}
    (T\psi_1)(x) = \sum_{c=0}^{\Nc-1} \psi_{1,c}\phi_c(x),
    \label{eq:prolongator}
\end{align}
where we suppress the spin and colour indices on the quark fields.
Note that these operators are non-square if $\Nc$ is smaller than the dimension
of the space of quark fields $12V/a^4$ which will always be the case in practice.
In addition, the linear operator $RT$ acts as the identity operator on the
coarse space, while $P=TR$ projects quark fields to the subspace spanned by
the basis fields.
The prolongator defined in ~\cref{eq:prolongator} is related to the restrictor as $T=R^\dagger$.

Next, we define the Hermitian Dirac operator on the coarse grid
\begin{align}
    Q_1 = RQT,\qquad Q=\gamma_5D,
\end{align}
which acts in the space of coarse fields.
$Q_1^{-1}$ appears in the Schur complement of the block $Q_1$ of $Q$ and forms
the basis of the deflation of the Dirac equation where it is often referred to
as the little Dirac operator.
From here, we define the decomposition of the quark propagator 
\begin{align}
    S = S_0 + S_1
\end{align}
where the first term corresponds to the Green's function for the deflated
system
\begin{align}
    S_0 = S - TQ^{-1}_1R\gamma_5,
\end{align}
while the second term is the subspace contribution
\begin{align} S_1 =
    TQ^{-1}_1R\gamma_5,
\end{align}
and this decomposition can be inserted into any Wick contraction.
The inversion of the coarse grid operator is expected to be
computationally inexpensive, since the coarse operator dimension is much smaller than the full Dirac operator.
We demonstrate that, if the coarse operator is properly constructed, most of the additional fluctuations due to the stochastic estimator of the translation average are captured in the coarse space, and thus can be easily suppressed.  
In \cref{sec:lma} we review the basic LMA method applied to the isovector vector current correlator, and set the notation for the multigrid LMA discussed in \cref{sec:loccoh} and \cref{sec:mglma}.

\subsection{Low-mode averaging for the current correlator}
\label{sec:lma}

In this section we define a simple scheme applied to the bare isovector vector
current correlator in the time-momentum representation \cite{Bernecker_2011}, which is defined by
the expectation value of the Wick contraction
\begin{align} \label{eq:tmr}
    C(x_0,y_0) &= -\frac{a^6}{3L^3}\sum_{k=1}^3
    \sum_{\bm x,\bm y}\tr\{S(x,y)\gamma_k S(y,x)\gamma_k\}
\end{align}
where we have assumed the local discretization of the current, and ignored any
improvement counterterms which may be required to remove $\mathrm O(a)$
discretization effects.
By plugging in the decomposition for the quark propagator, we arrive at the
decomposition of the correlation function in two terms
\begin{align}
    C(x_0,y_0) &= C_{11}(x_0,y_0) + \{C(x_0,y_0)- C_{11}(x_0,y_0)\}
\end{align}
where the first term contains only contributions from the coarse operator
\begin{align} \label{eq:lma:ee}
    C_{11}(x_0,y_0) = -\frac{a^6}{3 L^3} \sum_{k=1}^3 \sum_{\bm x,\bm y}\tr\{S_1(x,y)\gamma_k
    S_1(y,x)\gamma_k\}
\end{align}
while each term in the remainder
\begin{align}
    C(x_0,y_0)-C_{11}(x_0,y_0) &= -\frac{a^6}{3 L^3} \sum_{k=1}^3 \sum_{\bm x,\bm y}\Big[
        \tr\{S_0(x,y)\gamma_k S_0(y,x)\gamma_k\} \\
        & + \tr\{S_1(x,y)\gamma_k S_0(y,x)\gamma_k\}
          + \tr\{S_0(x,y)\gamma_k S_1(y,x)\gamma_k\}
      \Big],
\end{align}
involves at least one full quark propagator contained in $S_0$.

In low-mode averaging, the vectors $\phi_c(x)$ are chosen for example to be
$\Nc$ eigenvectors of $Q$ with smallest magnitude eigenvalues $\mu_c$, so the
operator $TQ^{-1}_1R$ coincides with the (truncated) spectral decomposition
\begin{align}
    S_1(x,y) &= \sum_{c=0}^{\Nc-1}\frac{1}{\mu_c} \phi_c(x)\phi_c^\dagger(y)\gamma_5
\end{align}
in which case the coarse operator $Q_1$ is diagonal and the trace in
$C_{11}(x_0,y_0)$ can be written in terms of the inner products
\begin{align} \label{eq:coarse:gamma}
    (\gamma_5 \gamma_k)_1(x_0)_{cd} = \sum_{\bm x} \phi^\dagger_c(x)\gamma_5\gamma_k\phi_d(x),
\end{align}
which can be evaluated with practically no further computational cost once the
basis fields have been determined.

On the other hand, the terms in the remainder are costly to evaluate with full
translation averaging as each term involves at least one full quark
propagator.
It is clear then, that we should not need an exact evaluation of the remainder
term.
A precise enough estimate could be obtained, for example, if \emph{(i)} it is
estimated with a simple stochastic estimator, such as the usual one-end
trick~\cite{Michael:1998sg,hutch_1990} and if \emph{(ii)} the low-mode
subspace is large enough so the variance on this term is sufficiently
suppressed.
Practically, however, it is often observed that the low-mode subspace is too
small to suppress the variance sufficiently on the remaining term and then
further variance reduction techniques need to be employed, see
Ref.~\cite{Djukanovic:2024cmq} for a discussion.
As the volume is increased with a fixed number of low modes used, this problem
is certain to deteriorate as the effect of the deflation becomes quickly
negligible.
Various alternative variance reduction methods must then be employed to
suppress the variance on this term in different
ways~\cite{Bali_2009,Blum_2013,Kuberski_2023}.
In this work we follow a different strategy which directly increases the
dimension of the low-mode subspace by exploiting the local coherence or weak
approximation property of the low quark modes, as discussed in the following
\cref{sec:loccoh}.

\begin{table}[t]
    \centering
    \begin{tabular}{ccccc}
        \toprule
        {ensemble}&
        {$L_0 \times L^3$}&
        {$L$ ($\mathrm{fm}$)}&
        {$m_{\pi} L$}&
        {\# configs} \\
        \midrule
        E7              & $64  \times 32^3$ & 2.1 & 2.9 & 100 \\
        F7 & $96  \times 48^3$ & 3.2 & 4.3 & 100 \\
        G7              & $128 \times 64^3$ & 4.2 & 5.8 & 100 \\
        \emph{(H7}              &
        $\mathit{192 \times 96^3}$ &
        \emph{6.3} &
        \emph{8.6} &
        \emph{5)}   \\
        \bottomrule
    \end{tabular}
    \caption{\label{tab:ensembles}%
        Ensembles used in this work.  All lattices have a pion mass
        $m_{\pi} = 270$ MeV and a lattice spacing of $a = 0.0658(10)$ fm
        with $\Nf=2$ $\mathrm O(a)$-improved Wilson fermions with lattice
        coupling $\beta=5.3$, hopping parameter $\kappa=0.13638$ and
        $c_\mathrm{sw} = 1.90952$~\cite{cls,Jansen:1998mx}. The largest H7 ensemble is
        only used to compute some variances which can be accurately determined
        from a few measurements.}
\end{table}

\section{Local coherence and block decomposition}
\label{sec:loccoh}

The observation of the local coherence or weak-approximation property which
states that in local regions of space-time, the low-mode subspace is spanned
by a small number of basis modes has been crucial in the development of
efficient preconditioning or deflation of the usual Krylov subspace solvers used
for the Dirac equation~\cite{Luescher2007,babich2010}.
In contrast to that case where inexact low modes are used to create the
deflation subspace, in the following we will consider only the case
where exact low modes have been determined%
\footnote{Throughout the document, we refer to exact modes
$\psi$ as modes with a residual of $r = \norm{Q \psi - \lambda \psi} / \norm{\psi}
\leq 10^{-12}$, where $\lambda$ is the corresponding eigenvalue.}, although the extension of the
method to using modes determined with a low accuracy is straightforward as we
will not assume any special properties of the basis fields.
This property has been exploited for the acceleration of the Lanczos algorithm
used to generate low modes~\cite{Clark_2018}.
In this section, we review the main idea and provide some numerical tests of
the concept using ensembles of $\Nf=2$ non-perturbatively $\mathrm
O(a)$-improved Wilson fermions.

We assume to have computed a small number $\Nc$ low quark modes
$\phi_0(x),\ldots,\phi_{\Nc-1}(x)$ of the Hermitian operator $Q$.
Now consider a regular block decomposition of the lattice
\begin{align}
    \label{eq:block_decomp}
    \Lambda &= \{ x = (x_0,x_1,x_2,x_3)\mid x_\mu \in \{0, \dots, L_\mu-a\} \}
    = \bigcup_{y \in \Lambda_1} B_y
\end{align}
into $V_1$ disjoint blocks $B_y$ of side length $b_\mu$ defined
\begin{align}
    \label{eq:block_def}
    \qquad B_y = \{ x\in\Lambda \mid x_\mu - y_\mu b_\mu < b_\mu\}
\end{align}
and labelled by the coordinate $y \in \Lambda_1$ in the block grid
\begin{align}
    \label{eq:block_grid}
    \Lambda_1 &= \{ y = (y_0, y_1, y_2, y_3) \mid y_\mu \in \{0, \dots,
L_\mu/b_\mu-1\}\},
\end{align}
where we assume $b_\mu$ divides the lattice size $L_\mu$.
Then we define the block-projected fields as
\begin{align}
    \label{eq:block_field}
    \psi_c^{B_y}(x) = \theta_{B_y}(x)\phi_c(x)
\end{align}
using the characteristic function on the block
\begin{align}
\theta_B(x) =
    \label{eq:char_func}
\begin{cases}
    1 & \mathrm{if}\quad x \in B,\\
    0 & \textrm{otherwise}.
\end{cases}
\end{align}
Such fields are no longer orthonormal and they may be orthonormalized again
locally on the blocks, which we denote by $\phi_c^{B_y}(x)$.

Local coherence states that the set of block fields provide a good description
of the actual low modes $\phi_c(x)$ with $c\geq\Nc$, so that the low modes are mostly
contained in the space spanned by the block fields with small deficits
\begin{align}
    \label{eq:cdeficit}
    \epsilon_c = \norm{(1-P)\phi_c}.
\end{align}
where $P$ is now understood to refer to the projection operator defined using
the set of block-projected fields $\phi^{B_y}_c(x)$ in the sense of
\cref{sec:deflation}.

\begin{figure}[t]
    \centering
\includegraphics[width=0.75\textwidth]{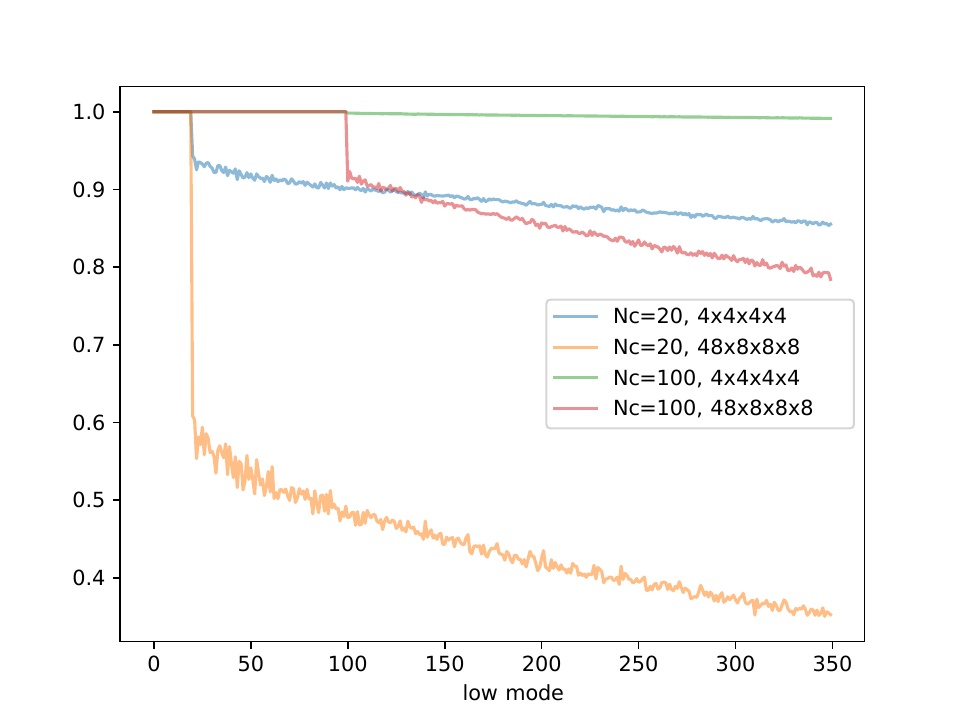}
\caption{\label{fig:lc} Local coherence of the low modes of $Q$ on ensemble
F7. The figure shows the value of $1-\epsilon_c$ versus the eigenmode
number $c$, where $\phi_c$ is the $c$-lowest mode for two different values of
$\Nc$ and block sizes.}
\end{figure}

To illustrate this, we computed the deficits for the lowest modes with
$c<350$ of $Q$ by block-projecting $\Nc=20$ and $100$ low quark
modes on a thermalized configuration from an ensemble with $\Nf=2$
non-perturbatively $\mathrm O(a)$-improved Wilson fermions with a pion mass
around $m_\pi\approx270\,\mathrm{MeV}$ and $m_\pi L\approx4$, labelled F7 in
Tab.~\ref{tab:ensembles}.
In Fig.~\ref{fig:lc} we show the results for spatial block sizes of $b/a=4$
and $8$ which corresponds in physical units to about $b\simeq
0.25\,\mathrm{fm}$ and $0.5\,\mathrm{fm}$ respectively.
The deficits for $c<\Nc$ vanish identically, and the deficits increase as
$c$ is increased meaning that the modes with larger mode number are not so
well represented in the space spanned by the block-projected low modes.
The smaller blocks and larger number of modes have very small deficits in the
range of low modes considered, but even the case $\Nc=20$ has small deficits
of just about 10\% over the range of modes considered.

Next, we use the
block-projected fields to define the coarse grid operator.
The dimension of the coarse grid subspace will now be multiplied by a factor
of the coarse lattice volume $\Nb$, making it substantially larger than in typical low-mode averaging scenarios, yet still smaller than the Dirac operator itself.
This implies that the coarse-grid contribution to the correlator will need to be
evaluated stochastically. Its variance can be effectively suppressed due to the reduced
cost of the coarse-grid inversion, while the variance on the remainder will be
strongly suppressed.
In \cref{sec:mglma}, we introduce the method in a hierarchical way by
leveraging the fact that a single set of $\Nc$ fields can be block
projected to different blocks. This provides a natural decomposition of the
quark propagator in a sequence of operators with different dimensions.

\section{Multigrid decompositions}
\label{sec:mglma}

In this section we introduce the decomposition of the quark propagator based
on block-projected low modes as discussed in the previous section.
We generalize the method immediately to a hierarchical $N_\ell$-level
scheme which naturally arises from nesting the block decomposition of the
lattice.
It is convenient to use a notation which emphasises the proposed projection
preserves some of the sparsity structure of the Dirac operator.
First we introduce the sequence of block decompositions which we employ,
followed by the associated restrictor and prolongator operators which define
the decomposition of the quark propagator.

\begin{figure}[t]
    \centering
\includegraphics[height=0.9\linewidth,angle=-90,origin=c]{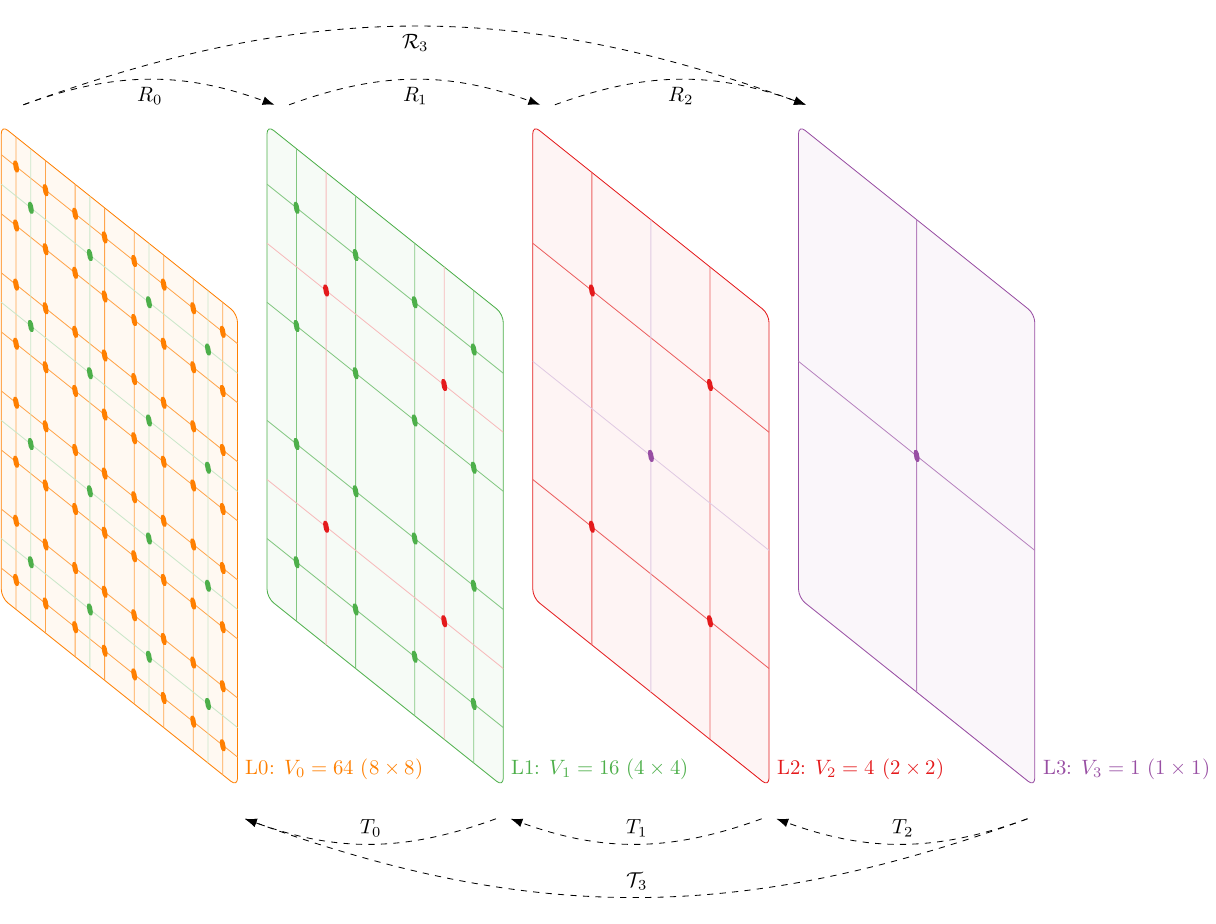}
\caption{\label{fig:nesting} 2D illustration of a recursive lattice coarsening via nested decompositions. The finest lattice on level 0 in orange has a size of $V_0/a^2=64$. The block size for level 1 in green is $b_0\times b_1=2a\times2a$ resulting in a coarse lattice size of $V_1=16$. The coarsest lattice in purple has a trivial lattice size of $V_3=1$ and thus corresponds to LMA, where no blocking is done. The restrictors $R_l$ and prolongators $T_l$ transfer data from one grid to the other as defined in the main text.}
\end{figure}

A hierarchical block decomposition scheme can be defined as follows, for
simplicity using regular decompositions into equal-sized blocks, which is a
requirement that can easily be relaxed later.
For each level $l=1,\ldots,\Nl-1$ we define a block decomposition of the
original lattice $\Lambda$ (sometimes referred to as the fine grid and denoted
$\Lambda_0$ for completeness) into $V_l$ disjoint blocks $B_y^l$ of side
length $b^l_\mu$ in complete analogy with the block decomposition in
\cref{sec:loccoh}.
To each decomposition we associate the block grid $\Lambda_l$ similar to
\cref{eq:block_grid}.
We impose then that the blocks of a given level $l$ also form a decomposition
of the blocks on the next level $l+1$, so that $b^{l+1}_\mu / b^{l}_\mu$ is a
natural number, which then clearly satisfy
\begin{align}
    B_z^{l+1} = \bigcup_{\{y\mid B_y^l\subset B_z^{l+1}\}} B_y^l.
    \label{eq:block_nested_decomp}
\end{align}
Such a scheme leads to a recursive definition for the coarse grid operators,
which, although not essential, is quite natural (see \cref{fig:nesting}).

For the outermost level, we define the restrictor $R_0: \vslattice_0\rightarrow
\vslattice_1$ and prolongator $T_0: \vslattice_1\rightarrow \vslattice_0$
similarly to \cref{sec:deflation}, between the space of quark fields of
dimension $12V/a^4$ to the space of coarse fields of dimension $\Ns\Nc V_1$ using
the basis fields constructed in the following way.
We explicitly set $\Nc$ to the number of low quark modes, but in contrast to
\cref{sec:loccoh} we set $\Ns=2$ and define the fields
\begin{align}
    \psi^{B^1_y}_{\alpha c}(x) = P_\alpha\theta_{B_y^1}(x)\phi_c(x),
    \qquad\alpha=0,1,\quad c=0,\ldots,\Nc-1,\quad y\in\Lambda_1
    \label{eq:mg_basis}
\end{align}
where $P_\alpha = \tfrac{1}{2}(1+(-1)^\alpha\gamma_5)$ are chiral projectors
and the colour and spin indices are suppressed as before.
The motivation to preserve this particular spin structure will be discussed in
\cref{sec:results} and \cref{appendix:chirality}.
Finally, we denote the new basis fields by $\phi_{\alpha c}^{B^1_y}(x)$ after
orthonormalizing those in \cref{eq:mg_basis} using, for example, Gram-Schmidt
orthogonalization.

The coarse fields $\psi_l\in\vslattice_l$ associated with the level $l>0$ of
the block decomposition are defined by the complex numbers $\psi_{l,\alpha
c}(y)$ for $\alpha=0,\ldots,\Ns-1$, $c=0,\ldots,\Nc-1$ and $y\in\Lambda_l$.
Analogously to \cref{sec:lma}, the corresponding operators are then defined
for $\psi_0\in\vslattice_0$ using the new basis fields via
\begin{align}
    \label{eq:res_mg}
    (R_0\psi_0)_{\alpha c}(y)
    =   \sum_{x\in\Lambda_0}(\phi^{B_y^1}_{\alpha c})^\dagger(x) \psi_0(x),
\end{align}
and for $\psi_1\in\mathcal V_1$
\begin{align}
    \label{eq:pro_mg}
    (T_0\psi_1)(x)
    =   \sum_{y\in\Lambda_1} \sum_{\alpha=0}^{\Ns-1}\sum_{c=0}^{\Nc-1}
    \psi_{1,\alpha c}(y) \phi^{B_y^1}_{\alpha c}(x).
\end{align}

Associated with the nested domain decomposition outlined above is the sequence
of restrictors $R_l: \vslattice_{l}\rightarrow \vslattice_{l+1}$ and
corresponding prolongators $T_l: \vslattice_{l+1}\rightarrow \vslattice_{l}$ for
$l=1,\ldots,N_\ell-2$ which map between coarse spaces of dimension
$N_\mathrm{s}N_\mathrm{c}V_l$ where $\Ns$ and $\Nc$ are now fixed as above and
$\vslattice_{l}$ is the coarse lattice size of the corresponding block
decomposition.
For all inner levels then the action of the restrictor on a coarse field
$\psi_{l} \in \vslattice_{l}$ is defined via
\begin{align}
    (R_l \psi_{l})_{\alpha c}(y) = \sqrt{\frac{V_{l+1}}{V_{l}}}
    \sum_{z \in \Lambda_{l}} \theta_{B^{l+1}_y}(x_l(z))\psi_{l,\alpha c}(z)
\end{align}
and the corresponding action of the prolongator on a field $\psi_{l+1} \in \vslattice_{l+1}$ as
\begin{align}
    (T_l \psi_{l+1})_{\alpha c}(z) = \sqrt{\frac{V_{l+1}}{V_{l}}}
    \sum_{y \in \Lambda_{l+1}} \theta_{B^{l+1}_y}(x_{l}(z)) \psi_{l+1,\alpha c}(y)
\end{align}
in terms of the coordinate map $x_l(z) = (b^l_0
z_0,b^l_1z_1,b^l_2z_2,b^l_3z_3)$ on level $l$.
Now we can define the coarse grid Hermitian Dirac operator on any level by the
recursion
\begin{align}
    Q_{l+1}   &= R_l Q_l T_l
\end{align}
where we set $Q_0=Q$.
For convenience we also use the compound restrictors $\mathcal R_l : \vslattice_0 \rightarrow \vslattice_{l}$ and prolongators $\mathcal T_l : \vslattice_{l} \rightarrow \vslattice_0$ for $l>0$
\begin{align}
    \mathcal R_l &=  R_{l-1}\cdots R_0, \qquad \mathcal T_l =  T_0\cdots T_{l-1},
\end{align}
and define $\mathcal R_0 = \mathcal T_0 = \mathbf{1}$, which are identical to
single operations which describe directly the mapping between the fine grid and
the coarse space associated with level $l$.

By iterating the identity
\begin{align}
    Q_{l}^{-1} &= \{Q_{l}^{-1} - T_l Q_{l+1}^{-1} R_l\} + T_l Q_{l+1}^{-1} R_l
\end{align}
we arrive at a telescoping sum decomposition of the quark propagator
\begin{align} \label{eq:mglma:decomp}
    S = S_0 + S_1 + S_2 + \ldots + S_{N_\ell-1}
\end{align}
into the differences for the outer levels $l=0,\ldots,\Nl-2$
\begin{align}
    S_l &= \mathcal T_{l}Q_l^{-1}\mathcal R_{l}\gamma_5 - \mathcal T_{l+1}
    Q_{l+1}^{-1}\mathcal R_{l+1}\gamma_5,
\end{align}
and the final contribution involving only the coarsest grid
\begin{align}
    S_{N_\ell-1} &= \mathcal T_{N_\ell-1}Q_{N_\ell-1}^{-1}\mathcal
    R_{N_\ell-1}\gamma_5.
\end{align}
The different terms in the telescoping sum could be constructed
without utilizing a recursion
allowing for the restrictor and prolongator to be defined independently of one another, which would provide greater flexibility for the method.
However, the current construction already offers a high degree of control over
the size and cost of the individual terms.

\begin{figure}[t]
    \centering
\includegraphics{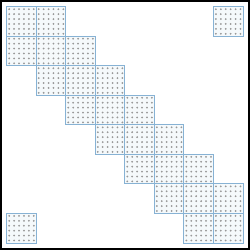}
\caption{\label{fig:A} 1D example with $V_{k} = 8$ of the structure of the
    coarse-grid operator $Q_k$ using periodic boundary conditions.
    Each block is of size $\Nc\Ns \times \Nc\Ns$.  Nearest neighbour
    interactions of $\gDop$ make neighbouring blocks of $Q_k$
    occupied shown in grey. Note that in 4D every block has 8 neighbours. The
    matrix has $(2 d+1)V_k \Nc^2\Ns^2$ non-zero entries, where
    $d$ is the dimensionality on space-time.}
\end{figure}

We are now equipped to explain how such a decomposition can be applied to compute
Wick contractions, such as the example of the isovector current correlator
discussed in \cref{sec:lma}.
As previously motivated, the block decomposition of the modes increases the effective size of the low mode subspace, particularly in the case of small block size, e.g. $8\times8^3$ or $4\times4^3$.
While the coarse-grid operators no longer assume a diagonal form as is typical in simplest low-mode averaging prescriptions, they retain a block
structure as illustrated in \cref{fig:A}.
The coarse and fine operators look similar in terms of sparsity, due to the blocking procedure. However, the coarse operator is usually less
sparse, because it contains more `colour' degrees of freedom $\Nc$.
In comparison to the fine operator, the coarse operators are also much smaller in magnitude,
which makes inversions less expensive and allows the implementation of efficient stochastic
estimators.
This argument about the efficiency, however, relies on the assumption that the coarse-grid operators are not only invertible
but also well conditioned.
In practice, we have observed that without retaining some of the spin
degrees of freedom on the coarse level as in \cref{eq:mg_basis}, the
coarse-grid operators can become ill conditioned.
With the outlined modification, we observe that the smallest eigenvalue of the
coarse operator is the same as that of the lattice Dirac operator on the same gauge
configuration, see \cref{appendix:spectra} for further details.

\subsection{Estimators for the isovector vector correlator}

As an instructive example, we apply the multigrid LMA method discussed in \cref{sec:mglma} to the bare isovector vector
current correlator in the time-momentum representation given in~\cref{eq:tmr}. 
We apply the decomposition of the quark propagator into $\Nl$ contributions,
\cref{eq:mglma:decomp}, and find the Wick contraction decomposes into terms
\begin{align} \label{eq:C_ij}
C_{ij}(x_0,y_0) &= -\frac{a^6}{3L^3}\sum_{k=1}^3
    \sum_{\bm x,\bm y}\tr\{S_i(x,y)\gamma_k S_j(y,x)\gamma_k\},\qquad i,j \in \{0,1, \dots,\Nl-1\}.
\end{align}
We recover \cref{eq:tmr} by summing $C (x_0,y_0) = \sum_{i,j} C_{ij}(x_0,y_0)$.
\begin{figure}[t]
\centering
\includegraphics{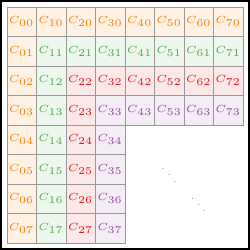}
\caption{\label{fig:C} Assignments of matrix elements of the correlator matrix $C_{ij}$, \cref{eq:C_ij}, to grid levels $\mathrm L k$. The sum of the orange shaded elements correspond to the $\mathrm L 0$-contribution of the correlator, green to $\mathrm L 1$, red to $\mathrm L 2$ etc.}
\end{figure}
We may further define the level-$k$-contribution to the full correlator as all components of the correlator matrix where either ($i=k$ and $j \geq k$) or ($j=k$ and $i \geq k$), i.e.
\begin{align}
C_{\mathrm L k}(x_0, y_0) = C_{kk}(x_0, y_0) + \sum_{i=k+1}^{\Nl-1} \Big( C_{ik}(x_0, y_0) + C_{ki}(x_0, y_0) \Big),
\end{align}
as illustrated in \cref{fig:C}, such that the full correlator can be recovered by summing over all levels,
\begin{align} \label{eq:cdecomp}
C(x_0, y_0) = \sum_{k=0}^{N_l-1} C_{\mathrm L k}(x_0, y_0).
\end{align}
One can choose different evaluation strategies for each of the level-contributions $C_{\mathrm L k}$.
A key observation is that evaluating $C_{\mathrm L k}$ requires inversions of the Dirac operator $Q_k$ on level $k$ and coarser, but not the inversions on finer levels $0,\dots,k-1$.
Level $\mathrm L0$ requires the computation of the full quark propagator.
The nearest-neighbour structure of $Q_k$ allows the usage of the same solver algorithms as for $\gDop$.
Finally, we also define the correlator averaged over the temporal extent of
the lattice, assuming periodic boundary conditions also in time
\begin{align}
    \label{eq:timetrans}
    G(t) &= \frac{a}{L_0} \sum_{y_0=0}^{L_0-a} C(t + y_0, y_0)
\end{align}
and the averages $G_{\mathrm Lk}(t)$, $G_{ij}(t)$ corresponding to the
level-$k$ terms and the individual terms in the decomposition, respectively.

\subsubsection{Two-level multigrid LMA}
In the case of two levels of multigrid, i.e. $\prop = \prop_0 + \prop_1$, 
we have decomposed the full correlator $C(x_0,y_0)$ into two contributions, $C(x_0,y_0) = C_{\mathrm L 0}(x_0,y_0) + C_{\mathrm L 1}(x_0,y_0)$. We will choose different evaluation strategies for each of the two terms.
On one hand, every evaluation of the $\mathrm L 0$-term requires to compute the full propagator $\prop$ by solving $\gDop \psi = \eta$ for some sources $\eta$.
On the other hand, every evaluation of the $\mathrm L 1$-term, we only need to solve the coarse-grid Dirac equation $\gDop_1 \psi_1 = \eta_1$ for some sources $\eta_1 \in \vslattice_1$.
This is significantly cheaper than solving for the full propagator on the fine lattice if the subspace was chosen such that its dimension is smaller than the fine lattice dimension.

Thus the strategy is the following: evaluate the $\mathrm L 0$-term using just
a few, say $1 \text{--} 10$ stochastic or point sources, whereas the $\mathrm
L 1$-term is evaluated with more, say $100 \text{--} 1000$ sources, as
required to suppress the corresponding variance.
Every level can be treated separately in terms of its evaluation strategy,
i.e.~number and type of sources. Sometimes it may be possible to compute the
coarsest level $k=N_{\ell}-1$ exactly, notice this is only beneficial if its dimension $\dim_{\mathbb{C}}(\vslattice_k) = N_c N_s V_k$ is small enough.

\subsubsection{Exact estimator on the coarsest grid}
If the coarsest subspace dimension is not too large (say $\dim_{\mathbb{C}}(\vslattice_{N_{\ell}-1}) < 10^4$), the coarsest term can be evaluated exactly if brought into the form
\begin{align} \label{eq:G_LL:exact}
G_{\mathrm L (N_{\ell}-1)}(t)
&= -\frac{a^7}{3 L_0 L^3} \sum_{k=1}^3 \sum_{y_0=0}^{L_0-a}
\tr
\left\{ (\gamma_5 \gamma_k)_{N_{\ell}-1}(y_0+t) (\gDop_{N_{\ell}-1})^{-1} (\gamma_5 \gamma_k)_{N_{\ell}-1}(y_0) (\gDop_{N_{\ell}-1})^{-1} \right\},
\end{align}
where the $(\gamma_5 \gamma_k)_{N_{\ell}-1}(x_0)$ are defined analogously as in \cref{eq:coarse:gamma} with an open (unsummed) index $x_0$.
When comparing this to the full lattice volume-averaged Wick contraction
given by \cref{eq:tmr,eq:timetrans}
we observe some similarities in the index structure thanks to the preservation
of the sparsity of the Dirac operator.
The trace is now over the coarse lattice indices, we have the coarse
propagator and Dirac matrices instead of the fine ones. Additionally the
dependence of the spacetime points $x,y \in \ispacetime$ is factorized, making
a full lattice volume average straightforward.
Since the term is calculated exactly and averaged over the full lattice
volume, the remaining variance will by definition be the gauge variance of
level $N_{\ell}-1$.
Furthermore, if $N_{\ell}=2$ and we use the exact low modes $\phi_c$ of $Q$,
the coarse lattice is trivial and $\Ns=1$, we will find the so-called
eigen-eigen term of LMA \cref{eq:lma:ee}.

\begin{table}
    \centering
    \begin{tabular}{cccccc}
    \toprule
             &  &       & \multicolumn{3}{c}{block size $b_0/a\times(b_j/a)^3$} \\
    \cmidrule(lr){4-6}
    ensemble &  estimator & $\Nl$ & $l=1$      & 2 & 3 \\
    \midrule
    E7 & LMA          & 2 & $64\times32^3$  & -              & -               \\
       & 2-lvl MG LMA & 2 & $8\times8^3$    & -              & -               \\
       & 3-lvl MG LMA & 3 & $8\times8^3$    & $64\times32^3$ & -               \\
    \midrule
    F7 & LMA          & 2 & $96\times48^3$  & -              & -               \\
       & 2-lvl MG LMA & 2 & $8\times8^3$    & -              & -               \\
       & 3-lvl MG LMA & 3 & $8\times8^3$    & $96\times48^3$ & -               \\
    \midrule
    G7 & LMA          & 2 & $128\times64^3$ & -              & -               \\
       & 2-lvl MG LMA & 2 & $8\times8^3$    & -              & -               \\
       & 4-lvl MG LMA & 4 & $4\times4^3$    & $8\times8^3$   & $128\times64^3$ \\
    \midrule
    H7 & LMA          & 2 & $192\times96^3$ & -              & -               \\
       & 2-lvl MG LMA & 2 & $8\times8^3$    & -              & -               \\
       & 4-lvl MG LMA & 4 & $4\times4^3$    & $8\times8^3$   & $192\times96^3$ \\
    \bottomrule
    \end{tabular}
    \caption{Block decompositions for the three estimators used in this work and the corresponding coarse-grid operators on levels $l$. When the block size matches the lattice volume, as is the case in simplest LMA schemes, no blocking is performed on that level. The coarsest level is not required to avoid
        blocking; for example the $\Nl=2$ MG LMA schemes terminate on a
        block-projected level.}
    \label{tab:blocks}
\end{table}

\subsubsection{Stochastic estimator on level \texorpdfstring{$k$}{k}}
\label{sec:stochest}

Every term in the decomposition \cref{eq:cdecomp} can be evaluated differently.
To estimate one term stochastically with the Hutchinson method
\cite{hutch_1990,Michael:1998sg}, we use the same noise fields as in the fine grid case.
With $N_{\eta}$ stochastic time-diluted fine-grid wall-sources
$\eta_{n}^{(t_n)}$ with support on time-slice $t_n$
chosen uniformly from $\{0, \dots, L_0/a - 1\}$
\begin{align}
\eta_{n}^{(t_n)}(x) &= 0 \qquad \text{ unless } \quad t_n = x_0
\end{align}
and having the properties of zero mean 
\begin{align}
    \lim_{\Nsrc\rightarrow\infty}\frac{1}{N_{\eta}} \sum_{n=0}^{N_{\eta}-1} \eta_{n}^{(t_n)}(x)_{\alpha a} &= 0, 
\end{align}
and unit variance
\begin{align}
\lim_{N_{\eta} \to \infty} \frac{1}{N_{\eta}} \sum_{n=0}^{N_{\eta}-1}
\eta_{n}^{(t_n)}(x)_{\alpha a} \left(
\eta_{n}^{(t_n)}\right)^\dagger(y)_{\beta b} &= \delta_{\alpha \beta}
\delta_{ab} \delta_{xy}.
\end{align}
the estimator for the correlator on levels $i,j$ is then
\begin{align} \label{eq:C_stoch}
    \mathcal G_{ij}(t) &= \frac{1}{N_{\eta}}
    \sum_{n=0}^{N_{\eta}-1}\mathcal G_{ij}^n(t_n+t,t_n),
\end{align}
where
\begin{align} \label{eq:C_stoch2}
    \mathcal G_{ij}^n(x_0,y_0) &= -\frac{a^6}{3 L^3} \sum_{k=1}^3 \sum_{\bm x} \innerprod{
    (\prop_j \gamma_5 \gamma_k^\dagger \eta_n^{(y_0)})(x)
}{
    \gamma_5 \gamma_k
    (\propi{i} \eta_n^{(y_0)})(x)
}.
\end{align}
In this work, we chose random noise uniformly from the set $\{ 1+i,1-i,-1+i,-1-i \}/\sqrt{2}$, but choosing samples from any
(sub-)Gaussian distribution is sufficient.
The product $\propi{j} \eta$ is obtained by solving $\gDop_j \psi_j = \eta_j$ and $\gDop_{j+1} \psi_{j+1} = \eta_{j+1}$, where $\gDop_k$ is the coarse-grid Dirac operator on level $k$ and $\eta_k = \mathcal R_{k} \eta$ the restricted fine grid source.
We have chosen the spin-diagonal variant~\cite{ETM:2008zte} which allows
the unbiased estimation of each spin matrix element with just $4 N_{\eta}$
inversions.
The estimator for the level-$k$ term is given by fixing the same $\Nsrc$
source fields in each of the terms on the right hand side
\begin{align}
    \mathcal G_{\mathrm Lk}(t) = \mathcal G_{kk}(t) + \sum_{i=k+1}^{\Nl-1}
    \Big(\mathcal G_{ik}(t) + \mathcal G_{ki}(t)\Big).
\end{align}
We note that if the noise sources were fixed to be the same for
each term in the estimator $\mathcal G=\sum_{k}\mathcal G_{\mathrm Lk}$,  it would reduce
to the standard one-end trick estimator \cite{Michael:1998sg}.
In our application, we will use a different number of sources on
each level.
  
\section{Numerical study with \texorpdfstring{$\mathrm O(a)$}{O(a)}-improved Wilson fermions}

\label{sec:results}

In the following, we investigate how the implementation of various multigrid
decompositions works out in practice for the case of the isovector vector
correlator, both in terms of the contributions to the variance and the computational cost.
In particular, we study the variance as a function of the physical size
of the lattice, spanning a factor two in $L$ from $m_\pi L=2$ to $6$.
We use $\Nf=2$ flavours of non-perturbatively $\mathrm O(a)$-improved Wilson
fermions all with the same bare lattice parameters of the CLS ensemble
labelled by F7 \cite{cls} in \cref{tab:ensembles}.
The corresponding pion mass on that ensemble is about
$m_{\pi}\approx270\,\mathrm{MeV}$ with lattice spacing
$a=0.0658\,\mathrm{fm}$.
The F7 ensemble was generated by the CLS consortium~\cite{cls}, while we generated the
other ensembles with smaller and larger lattice sizes using the \texttt{openQCD-2.4}~\cite{openqcd} package.

\begin{figure*}[t]
    \centering
    \includegraphics[width=1.0\linewidth]{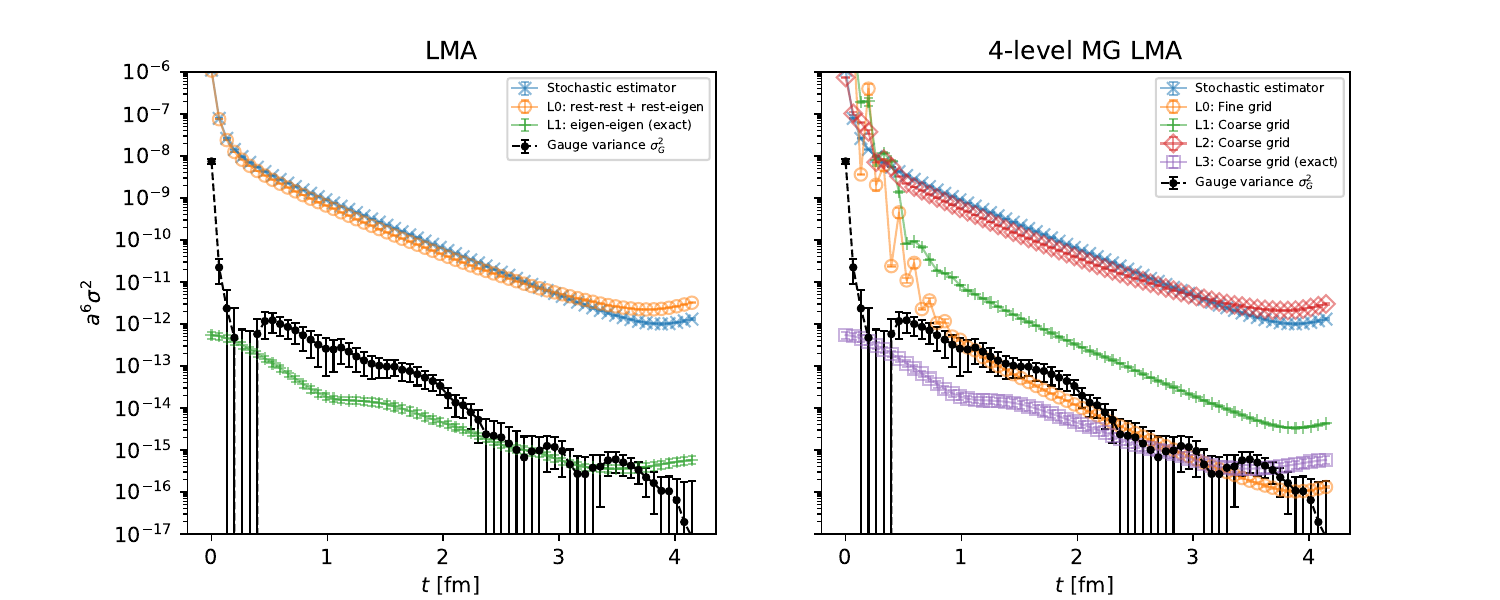}
    \caption{Absolute variances of different contributions to the the full vector correlator defined in \cref{eq:cdecomp} computed on the gauge ensemble G7 for LMA (left panel) and MG LMA (right panel) with one stochastic source for each term. The black filled circles denote the gauge variance. The standard stochastic estimator (blue crosses) is shown on both panels for reference.}
    \label{fig:G7:var:abs}
\end{figure*} 

For each ensemble, we evaluated three types of stochastic estimators for the
translation-averaged isovector vector correlator as defined in \cref{sec:stochest}: \emph{(i)} a plain stochastic estimator without any deflation of the low
modes, i.e.~the one-end trick, \emph{(ii)} low-mode averaging without any block projection of
the low modes which we refer to as LMA and \emph{(iii)} multigrid low-mode
averaging (MG LMA) with at least one level of block-projected modes.
If low modes without block projection are used, the coarse contribution
is also evaluated exactly.
For all ensembles, we present results with a fixed number of $\Nc=50$ low
modes and block sizes of approximately $0.25\,\mathrm{fm}$ and
$0.5\,\mathrm{fm}$.
The details of the block decompositions used for different gauge ensembles are given in
\cref{tab:blocks}.
To determine low modes of the Hermitian Dirac operator $Q$, we used the library \texttt{PRIMME} \cite{PRIMME}. 

In \cref{fig:G7:var:abs} we show the total variance in lattice units as a
function of the time-separation $t$ between the currents in physical units
on the G7 ensemble where $L\approx 4\,\mathrm{fm}$ for LMA (left panel) and MG LMA
(right panel).
The total variance for the plain stochastic with $\Nsrc=1$ source
is shown in both panels for comparison (blue crosses). 
The black circles depict the variance of the exact translation-averaged
correlator, or gauge variance, estimated using \cref{eq:appendix:gv:formula}, which decays
exponentially as expected from the well-known Parisi-Lepage
arguments~\cite{Parisi:1983ae,Lepage:1989hd}.

\begin{figure*}[t]
    \centering
    \includegraphics[width=1.0\linewidth]{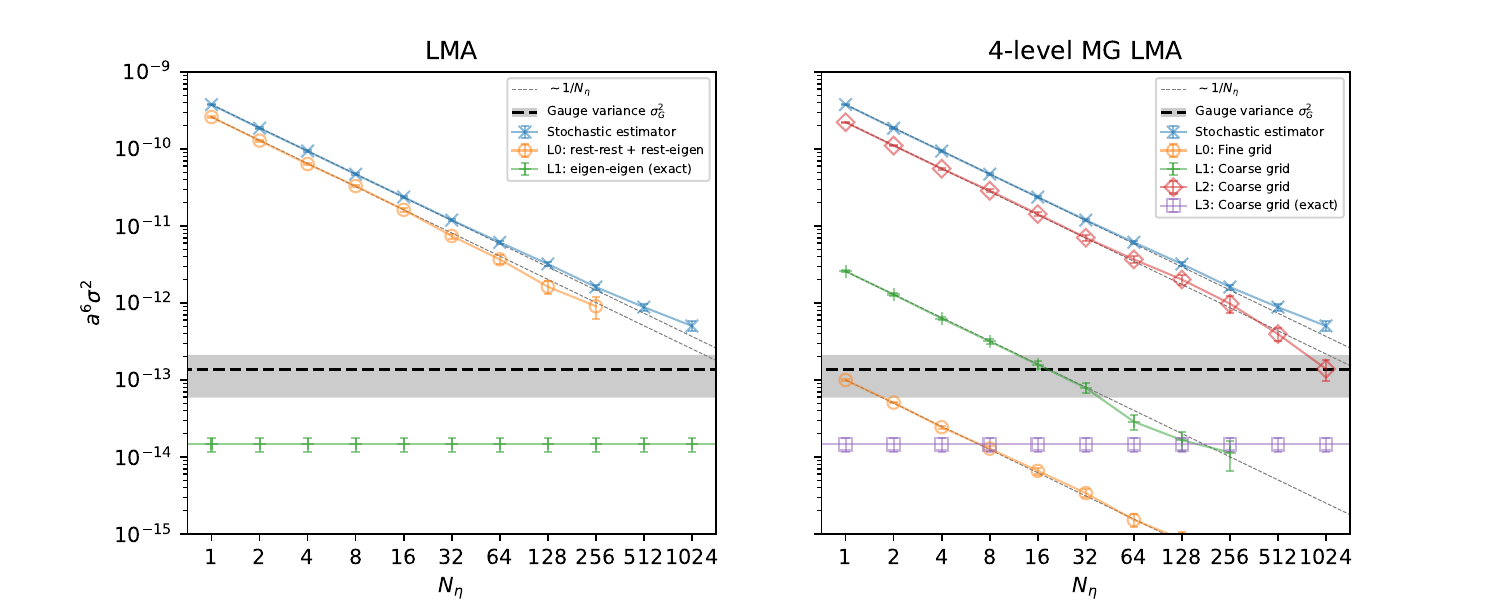}
    \caption{Absolute variances of the lattice G7 for LMA (left) and MG LMA
    (right) against the number of stochastic sources $\Nsrc$ on time
separation $t = 1.3\,\mathrm{fm}$. The black dashed line is the gauge variance. The standard stochastic estimator (blue crosses) appears in both plots for reference.}
    \label{fig:G7:var:sources}
\end{figure*}

For LMA the variance on the remainder terms labelled L0, which involve
inversions of the full lattice Dirac operator, are as large as the total
variance on the plain stochastic estimator. This indicates that a fixed number of low modes used here is insufficient to suppress the variance on the L0
terms for this physical volume.
In contrast, in the right-hand panel, for the MG LMA estimator based on the
same number of low modes, the variances show a different hierarchy, where
the stochastic variance on the coarser levels L1 and L2 is larger than the
fine level L0.
Given that the L2 operator has a dimension about 500 time smaller than the
lattice Dirac operator, we can leverage the reduced cost of the inversions to
suppress the variance on this term.

In order to estimate how many inversions are required to reach the target
gauge noise, we show the absolute variance at fixed separation
$t\approx1.3\,\mathrm{fm}$ on each contribution in
\cref{fig:G7:var:sources}, where the grey band indicates the gauge variance.
From here, we can see that the variance on the finest level needs just one
source to reach the target, while the next coarsest needs approximately
$\Nsrc\approx16$ and so on for the MG LMA.
Alternative choices for the time separation $t$ do not change the picture
qualitatively.

\begin{figure*}[t]
    \centering
    \includegraphics[width=1.0\linewidth]{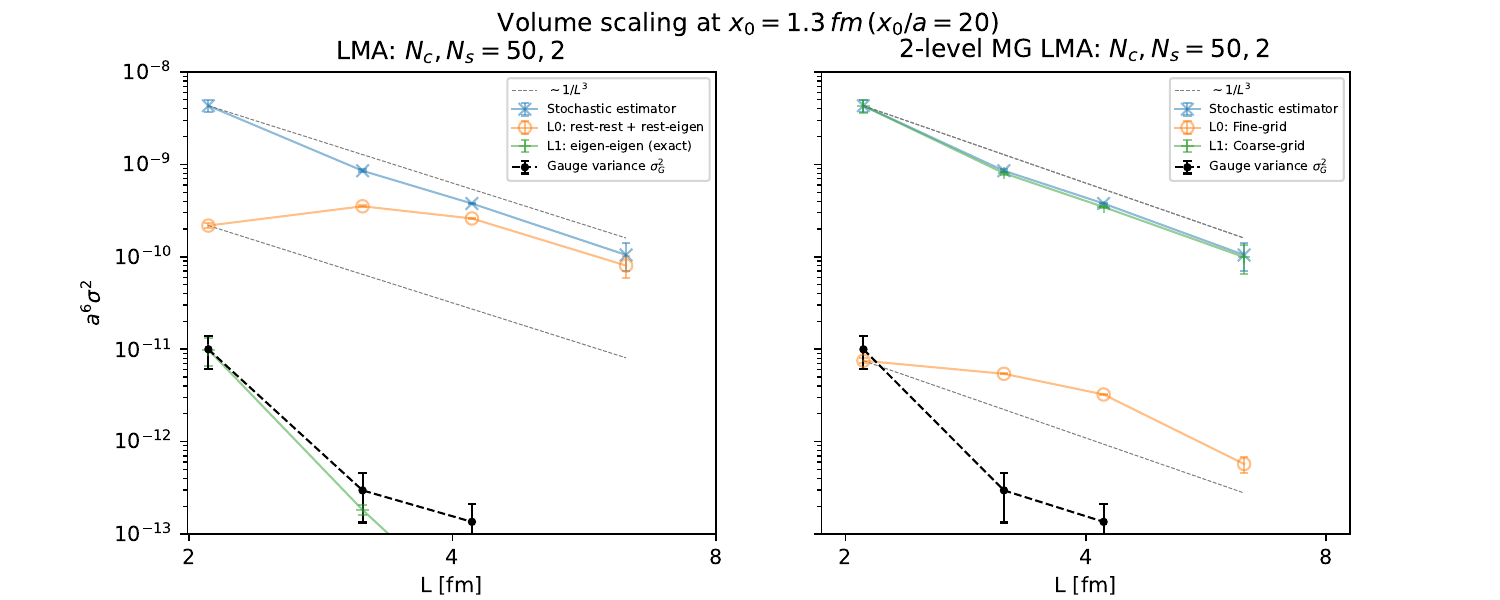}
    \caption{Absolute variances for LMA (left) and MG LMA (right) against the lattice extent $L$ on a log-log scale with one stochastic source for each term. The black dashed line is the gauge variance. The standard stochastic estimator (blue crosses) appears in both plots for reference.}
    \label{fig:var:vol}
\end{figure*}

Finally, we repeat the computation of the variances for different lattice
sizes in \cref{fig:var:vol}.
In this case we show the results from the $\Nl=2$ level MG LMA, i.e.~using
only one block-projected level (see \cref{tab:blocks}).
For LMA (left panel), we see that the variance of the fine level contribution
increases as the lattice volume increases until the variance coincides with
the plain stochastic estimator and the effect of the deflation is negligible.
On the other hand, for the MG LMA scheme (right panel), the variance of the
fine level contribution decreases over the whole range of lattice sizes,
indicating that increasing dimension of the block-projected mode space
suppresses the variance on this contribution effectively and thus achieves a
fairly constant variance reduction.

\subsection{Comparison of the cost}

\begin{table}[b!]
\begin{tabular}{ccccccccc}
\toprule
\multirow{2}{*}{{ensemble}} &
\multirow{2}{*}{{estimator}} &
\multirow{2}{*}{$\Nc$} &
\multicolumn{4}{c}{$\Nsrc$} &
\multicolumn{2}{c}{cost} \\
\cmidrule(lr){4-7}
\cmidrule(lr){8-9}
&&& L0 & L1 & L2 & L3 & measured & modelled \\
\midrule
E7 & Stochastic   & -  & 1024 & -     & -     & -     & $4096$  & $4096$  \\
   & LMA          & 50 & 16   & exact & -     & -     & $64$    & $64$    \\
   & 2-lvl MG LMA & 50 & 1    & 1024  & -     & -     & $100.4$ & $12.3$  \\
   & 3-lvl MG LMA & 50 & 1    & 16    & exact & -     & $5.5$   & $4.1$   \\
\midrule
F7 & Stochastic   & -  & 2048 & -     & -     & -     & $8192$  & $8192$  \\
   & LMA          & 50 & 1024 & exact & -     & -     & $4096$  & $4096$  \\
   & 2-lvl MG LMA & 50 & 16   & 2048  & -     & -     & $462.3$ & $80.7$  \\
   & 3-lvl MG LMA & 50 & 16   & 1024  & exact & -     & $263.2$ & $72.3$  \\
\midrule
G7 & Stochastic   & -  & 4096 & -     & -     & -     & $16384$ & $16384$ \\
   & LMA          & 50 & 2048 & exact & -     & -     & $8192$  & $8192$  \\
   & 2-lvl MG LMA & 50 & 16   & 2048  & -     & -     & $557.8$ & $80.7$  \\
   & 4-lvl MG LMA & 50 & 1    & 16    & 1024  & exact & $466.7$ & $14.4$  \\
\bottomrule
\end{tabular}
\caption{\label{tab:cost}
Comparison of the number of stochastic sources used on each level for
all schemes and ensembles. The associated cost is given in units of the cost
of one fine grid inversion. For levels where no blocking is used, no
stochastic estimator is required and the associated cost is zero.}
\end{table}

Next, we examine the cost of different estimators discussed in Sec.~\ref{sec:results} and tabulate in \cref{tab:cost} the approximate
number of sources required to reach the gauge noise.
We have observed that the cost does not depend strongly on the choice of
time separation $t$ in the accessible range.
Nevertheless, the gauge variance is fairly poorly determined and as such they
should be taken as indicative only.
The unit of cost is in inversions of the full lattice Dirac operator.
We do not include the cost of determining the low modes, since a small number
of low modes can safely be stored and reused for other observables, thus their
cost is amortised quickly%
\footnote{However, one may want to add $100-200$ to the numbers to include their cost. This amounts a cost of $2-4$ fine grid inversions per low mode.}.

\Cref{tab:cost} shows two values for the cost: measured cost and modelled
cost.
The measured cost is obtained using performance data from our actual
implementation reported in \cref{tab:cost:imp}.
We used a SAP-preconditioned GCR solver with inexact
deflation~\cite{Luscher_2003,Luescher2007} to solve the fine grid Dirac.
This solver was taken from the openQ$^\star$D software package \cite{openqxd_paper,openqxd_code}
which is based on openQCD \cite{openqcd}.
The coarse grid inversions are performed using a GCR solver preconditioned
with the deflated even-odd preconditioned coarse Dirac operator.
We expect the performance of the coarse grid solver to be far from
optimal, which can be observed from the high iteration counts in
\cref{tab:cost:imp}.
Indeed, we even observe one case where the time to solution for the coarse
grid solve is higher than the fine grid solve.

We establish a performance model in \cref{appendix:cond,appendix:mrhs} based
on a theoretical implementation of a multiple right-hand side solver whose
asymptotic cost scales proportionally to the dimension of the operator and assume
this implementation to exist on all grids.
Furthermore the performance model assumes the same solver on all grids, which will result in
smaller iteration counts for coarse grids (see \cref{appendix:mrhs}).
This feature would result in lower iteration counts on coarse grids than what
we observe in our implementation.
The modelled cost on the rightmost column in \cref{tab:cost} reports this
theoretical expected speedup using the given performance model.

In both cases the cost was calculated using the following formula
\begin{align}
    \textrm{cost} = 4 \sum_{k=0}^{\Nl -1} \frac{\Nsrc(\mathrm
Lk)}{\ratio (\mathrm Lk)},
\end{align}
where $\Nl$ is the number of MG levels of the estimator and $\Nsrc(\mathrm
Lk)$ was fixed with the number of stochastic sources of level $\mathrm Lk$
from \cref{tab:cost}.
The factor $4$ appears, because we use spin-diagonal random wall-sources which
amounts $4$ inversions per source.
For the measured cost, $\ratio(\mathrm Lk)$ was substituted with the values
from \cref{tab:cost:imp}, whereas for the modelled cost,
$\ratio(\mathrm Lk) = 12V/a^4/\Ns\Nc V_k$ for levels $k>0$ and
$\ratio(\mathrm L0) = 1$ (see \cref{eq:model:ratio}).
The cost of the exact coarsest grid evaluation was neglected for all
estimators since it only requires the direct inversion of an operator with
dimension $\Ns\Nc=100$.

For both the measured and modelled costs, we observe significant improvements
compared to the plain stochastic estimator.
Although the modelled cost is smaller than the cost observed on the two larger volumes, F7 and G7, 
it indicates the potential improvements that could be
achieved with a performant multiple right-hand side solver applied across all available grids.
For the smallest volume used here, E7, LMA already provides a significant
improvement over the stochastic estimator.
Even in this case, the effect of introducing an additional block-projected level is
significant, resulting in an order of magnitude improvement.
For the larger volumes, it is clear that LMA with few low quark modes
is far from optimal.
Extrapolating the number of low modes  from E7 to achieve an equal variance contribution
would result in $250$ low modes on F7 lattice, $800$ on G7, and about $4050$ on H7.
Therefore, we do not expect the comparison reported here to reflect a state-of-the-art
computationally set-up.
What is remarkable, however, is the significant improvement achieved with the fixed small
number of modes by introducing an intermediate
block-projected level.
Thus, with only a relatively minor overhead of generating a small number of low
modes, an efficient low-mode averaging scheme can also be constructed for large
physical volumes.

\begin{table}[b!]
\begin{tabular}{ccccrrr}
\toprule
{ensemble} & 
{type} & 
{grid size} & 
{block size} &
{time (sec)} &
{$\ratio = \textrm{coarse/fine}$} &
{\# iteration} \\
\midrule
E7              & fine   & $64 \times 32^3$  &       & $5.32(3)$  & $1.0$       & $35.7(2)$    \\
                & coarse & $8 \times 4^3$    & $8^4$ & $0.125$    & $42.6(2)$   & $140.5(3)$   \\
\midrule                                                                        
F7              & fine   & $96 \times 48^3$  &       & $8.42(4)$  & $1.0$       & $43.77(15)$  \\
                & coarse & $12 \times 6^3$   & $8^4$ & $0.409(2)$ & $20.6(1)$   & $337.6(1.3)$ \\
\midrule                                                                        
G7              & fine   & $128 \times 64^3$ &       & $11.1(4)$  & $1.0$       & $46.53(23)$  \\
                & coarse & $32 \times 16^3$  & $4^4$ & $37(2)$    & $0.29(2)$   & $1417(22)$   \\
                & coarse & $16 \times 8^3$   & $8^4$ & $0.67(4)$  & $16.6(1.2)$ & $502(6)$     \\
\bottomrule
\end{tabular}
\caption{\label{tab:cost:imp}
Cost of inversion of the fine and coarse grid operators. For the fine grid a
very efficient implementation is available in contrast to our probably
suboptimal coarse grid solver. The relative cost to the fine grid inversion
$\ratio$ is used to the compute the cost in \cref{tab:cost}.}
\end{table}

\section{Conclusions}
\label{sec:conclusion}

In this work, we have proposed a concrete implementation of a multigrid
variance reduction scheme based on block-projected low modes of the Dirac
operator. The block-projected low modes of the Dirac operator are crucial for
the formulation of deflation and multigrid preconditioning for the Dirac
equation and have previously been applied to variance reduction in the trace
of the quark propagator, compression of low modes, and the
acceleration of their generation.
Here, we extend these techniques to the quark propagator at long distances,
which describes the propagation of flavour non-singlet states like baryons and
isovector states.

Block projection allows one to efficiently span the low-mode subspace, 
which contains most of the contribution to the variance as well as the signal
for such states at large separations.
The low-mode subspace is still small enough that the corresponding contribution can be
evaluated inexpensively using simple stochastic estimators, as the
associated operators are relatively cheap to apply in comparison to the full
lattice Dirac operator.
The most important finding of this work is the demonstration that the
suppression of the variance remains independent of the physical volume when the
block size is held fixed.
This implies that determining just a small number of low modes, in the range
$10-100$ in our setup, is sufficient to construct an efficient scheme for the
isovector vector correlator, enabling the implementation of full translation
averaging.

The same scheme can be applied to any quark propagator but the efficiency
will likely depend on the precise structure of the observable.
The potential improvements in Wick contractions which involve more quark
propagators, for example baryon correlators, remains to be understood.
A more thorough understanding of the variance from a theoretical perspective,
by employing the spectral decomposition would clearly be useful.
The family of proposed schemes relies naturally on efficient true multigrid
solvers, and thus synthesises perfectly with the fast development of such
preconditioners in lattice QCD.
In this work we have not explored the use of inexactly determined low modes to
define the coarse-grid operators which may further reduce the overhead of the
method.
We suspect more concrete statements can be made about the condition number of
the coarse operator, although our numerical experiments show that the proposed
scheme is free from issues.

An implementation of multigrid LMA is expected to be less straightforward for the staggered fermion discretization of lattice QCD~\cite{KogutSusskindBanks, Susskind, Sharatchandra}. This is due to the appearance of spurious low eigenvalues in the coarse Dirac operator spectra, as discussed in the work introducing the multigrid algorithm for the two-flavour Schwinger model~\cite{Brower_2018}.
We observed a similar phenomenon with Wilson fermion discretisation of QCD, when coarsening without preservation of chiral degrees of freedom, see \cref{appendix:spectra}.
An alternative formulation of the multigrid algorithm uses the K\"ahler-Dirac operator \cite{becher1982,bodwin1988}, whose coarse spectra are well-behaved in the two-flavour Schwinger model~\cite{Brower_2018}, and the same behaviour is expected to hold in QCD.
We thus expect the proposed multigrid LMA procedure for variance reduction in hadronic observables to be applicable to the K\"ahler-Dirac formulation of staggered QCD. 

Finally, this variance reduction technique allows one to efficiently compute
the translation average for quark propagators, which with standard Monte Carlo
sampling mostly still suffer from signal-to-noise ratio problems.
These problems are exponentially improved by two-level sampling schemes~\cite{Ce:2016idq,Ce:2016ajy}, and
the observation that most of the gauge noise is contained in the block-decomposed low-mode subspaces may be of help in designing efficient factorizations of the quark propagator or the preconditioning of the HMC itself.

\hspace{1cm}

\noindent
{\it Acknowledgements.}---We are grateful to Richard Brower, Carlton De~Tar, Aida El-Khadra, Leonardo Giusti,
Steven Gottlieb, Christoph Lehner, Martin L\"uscher, Simon Kuberski, and the RC* collaboration
members for insightful discussions.
We acknowledge the access to Piz Daint at the Swiss National Supercomputing
Centre, Switzerland under the ETHZ’s share with the project IDs c21, eth8,
go21, and s1196 and especially thank Jonathan Coles from CSCS for
support and thoughtful input to our work.
The support for the PASC 2021-2024 project ``Efficient QCD+QED Simulations with openQ*D software" is gratefully acknowledged.
The authors gratefully acknowledge the Gauss Centre for Supercomputing e.V. (www.gauss-centre.eu) for funding this project by providing computing time on the GCS Supercomputer JUWELS~\cite{JUWELS} at Jülich Supercomputing Centre (JSC).
We acknowledge the EuroHPC Joint Undertaking for awarding this project access to the EuroHPC supercomputer LUMI, hosted by CSC (Finland) and the LUMI consortium through a EuroHPC Regular Access call.

\appendix

\section{Stochastic estimator for the gauge variance}

\begin{figure}[t]
    \centering
    \includegraphics[width=0.7\linewidth]{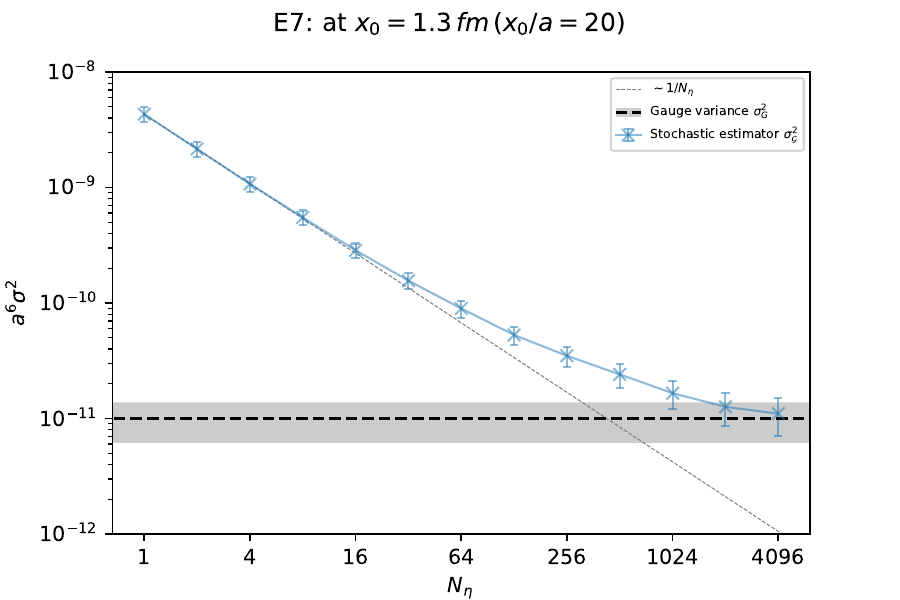}
    \caption{Gauge variance \cref{eq:appendix:gv:formula} (black dashed line)
    compared to the total variance as a function of $\Nsrc$ (blue crosses) of
\cref{eq:totvar} on lattice E7.}
    \label{fig:appendix:gv_reached}
\end{figure}

The variance of the Wick contraction $G(t)$ with exact translation averaging over
the full lattice, i.e.~the minimal or gauge variance, is given by
\begin{align}
\sigma^2_G(t) &= \langle G(t)^2 \rangle_U - \langle G(t) \rangle_U^2,
\end{align}
where $\langle\cdot\rangle_U$ represents the mean over the gauge fields.
We define the gauge variance such that the variance on the sample mean
over $N$ uncorrelated configurations would be $\sigma_G/\sqrt{N}$.
The stochastic samples can be used to estimate the gauge variance with the
following formula
\begin{align}
\nonumber
&\sigma^2_G \approx
\frac{1}{\Nsrc(\Nsrc-1)} \sum_{m \neq n=0}^{\Nsrc-1}\\
\label{eq:appendix:gv:formula}
\qquad&\Big[ \langle \mathcal G^m(y_0^m+t, y_0^m) \mathcal G^n(y_0^n+t, y_0^n) \rangle_U
- \langle \mathcal G^m(y_0^m+t,y_0^m) \rangle_U \langle \mathcal G^n(y_0^n+t,y_0^n) \rangle_U \Big] 
\end{align}
where $\Nsrc$ is the number of stochastic sources.
On the other hand, when a stochastic estimator $\mathcal G$ is used such as
the example given in \cref{sec:stochest}, the total variance includes a
contribution from the fluctuations of the auxiliary fields, and is defined as
\begin{align}
    \sigma^2_{\mathcal G}(t) &= \langle \mathcal G(t)^2 \rangle_U - \langle
    \mathcal G(t) \rangle_U^2,
    \label{eq:totvar}
\end{align}
which depends on $\Nsrc$.
Clearly, we have that $\lim_{\Nsrc\rightarrow\infty}\sigma^2_{\mathcal
G}=\sigma^2_G$.
In \cref{fig:appendix:gv_reached} we verify the limit is reached accordingly
for the isovector vector correlator at $t=1.3\,\mathrm{fm}$ on the E7
ensemble.

\section{Performance model}

\label{appendix:mrhs}

In \cref{appendix:cond}, we see that the condition number and consequently the iteration count of a Krylov solver decreases with coarser Dirac operators. Since the coarse Dirac operator is less sparse than the fine, it might happen in rare cases that its memory footprint is high -- even higher than the one from the fine grid Dirac operator. This introduces a problem, since inversions using Krylov solvers are memory-bound problems. We may introduce a performance model for the \df{cost of an inversion} of a sparse linear operator $K$ as
\begin{align}
\cost{K} = \iter{K} \big[ \mem{K} + 2 \cdot \mem{\psi} \big]
\end{align}
where $\iter{K}$ denotes the \df{number of iterations} required to invert $K$ for one right-hand side (RHS) using some Krylov solver and $\mem{K}$, $\mem{\psi}$ is the \df{memory traffic} produced by reading the operator $K$ and the spinor $\psi$ from memory.
We neglect scalar-products, norms, and axpy-type kernels in the iteration step, since the operator application dominates anyway.
In this scheme, the unit of cost is memory traffic in bytes which is adequate for memory-bound problems.
If we have a multiple-RHS capable solver available, we find the \df{cost for $N_{rhs}$ simultaneous solves} as
\begin{align}
\cost{K, N_{rhs}} = \iter{K} \big[ \mem{K} + 2 N_{rhs} \cdot \mem{\psi} \big].
\end{align}
Finally, we define the \df{speedup} of a coarse grid solve of $K_l$ on level $l>0$
with respect to a fine grid solve of $K$ in this model as the dimensionless quantity
\begin{align}
\mathcal S_l(N_{rhs}) = \frac{\cost{K, N_{rhs}}}{\cost{K_l, N_{rhs}}}.
\end{align}
In the asymptotic limit $N_{rhs} \to \infty$, we obtain
\begin{align}
\mathcal S_l(\infty) = \frac{\iter{K}}{\iter{K_l}} \frac{\mem{\psi}}{\mem{\psi_l}}.
\end{align}
Notice that $\mem{K}$ and $\mem{K_l}$ dropped out from the expression, meaning that the asymptotic speedup is independent of the size of the operators in memory. 
The speedup $\mathcal S_l(\infty)$ is then just the ratio of the dimensions of
the vector spaces.
This implies that $\mathcal S_l(\infty) \gg 1$ with many right-hand sides even if the coarse operator is large in memory.

In our performance model, we assumed conservatively that the iteration counts do not change $\iter{K} = \iter{K_l}$, even though \cref{appendix:cond} suggests $\iter{K} > \iter{K_l}$, and thus the speedup is given by the dimension ratio of the subspaces,
\begin{align} \label{eq:model:ratio}
\mathcal S_l = \ratio(\mathrm Ll) = \frac{12 V/a^4}{N_s N_c V_l}, 
\end{align}
where $V_l$ is the coarse lattice volume on level $l$.

Additionally, coarse grid Dirac operators are especially amenable to multiple RHS solvers, since coarse grid spinors usually have significantly smaller dimensions than fine grid spinors, $\mem{\psi} \gg \mem{\psi_l}$, thus with the same amount of memory, we can store more coarse fields than fine fields, i.e.~larger $N_{rhs}$.

\section{Chirality on the coarse subspaces}

\label{appendix:chirality}

One might ask the question, why not coarsen the Dirac operator $D$ itself as
is usually done in multigrid literature. The Dirac operator is not normal in
the Wilson formulation, thus we should consider singular vectors, rather than
eigenvectors as put forward in Refs.~\cite{Frommer:2013, babich2010}. The right
singular vectors of the Dirac operator are eigenmodes of the Hermitian Dirac
operator $Q$.

\begin{figure}[t]
    \centering
    \includegraphics[width=1.0\linewidth]{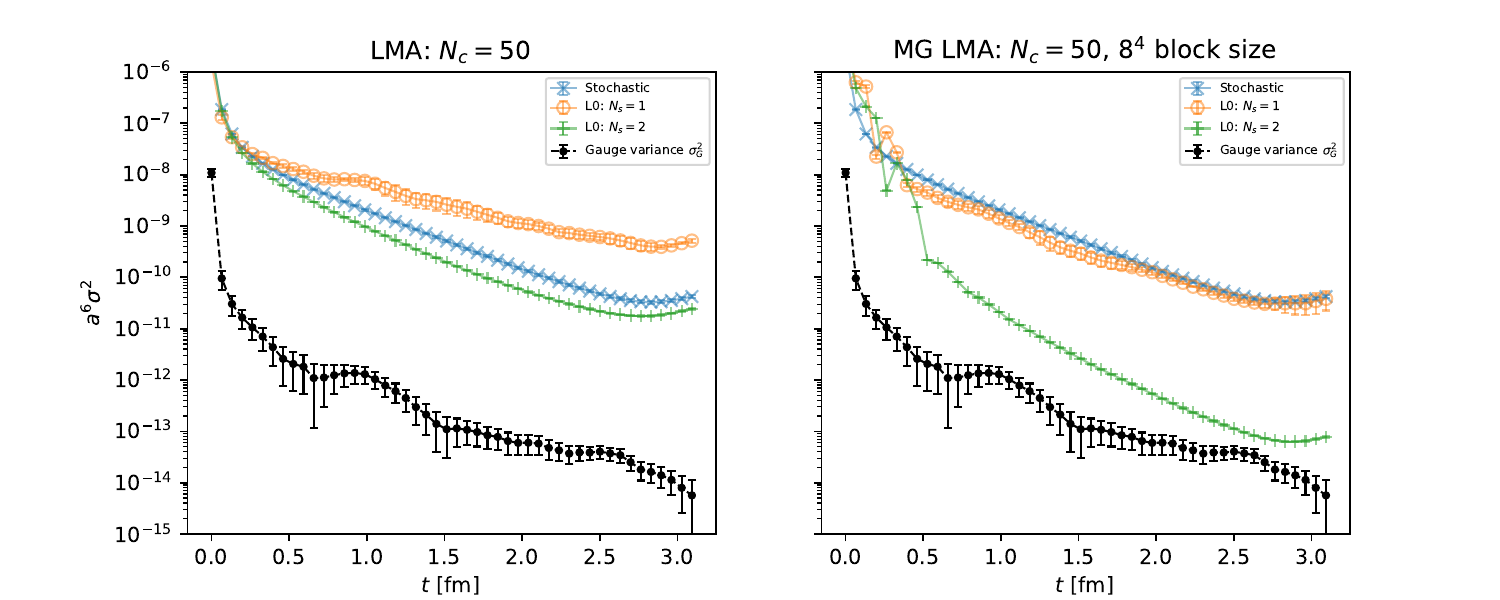}
    \caption{Impact of chirality preservation on the absolute variance of the L0-term for LMA (left) and $2$-level MG LMA (right) on lattice F7, when coarsening $D$ instead of $Q$. The yellow line shows the absolute variance when all the spin degrees of freedom are coarsened $N_s=1$, and the green line shows the twice-as-large subspace constructed with preservation of chirality $N_s=2$ as in \cref{eq:chirality:B}. The black dashed line is the gauge variance and the standard stochastic estimator (blue crosses) appears in both plots for reference.}
    \label{fig:chirality}
\end{figure}

Given the spectral decomposition of $Q$,
\begin{align}
Q = \sum_i \lambda_i \phi_i \phi_i^\dagger
\end{align}
we can construct the singular value decomposition of $D$,
\begin{align}
D = \sum_i \lvert \lambda_i \rvert \tilde{\phi}_i \phi_i^\dagger
\end{align}
where $\tilde{\phi}_i = \text{sign}(\lambda_i) \gamma_5 \phi_i$. We observe that left and right singular vectors of $D$ are proportional to $\gamma_5 \phi_i$ and $\phi_i$, respectively. The former are eigenmodes of $D \gamma_5$, whereas the latter are eigenmodes of $Q = \gamma_5 D$. This suggests that we should restrict using left singular vectors of $D$ and prolongate using the right ones. This would result in a Petrov-Galerkin coarsening, where $\mathcal{R}_l = \mathcal{T}_l^\dagger \gamma_5$, i.e. the coarse grid operator would be
\begin{align} \label{eq:PG}
\mathcal{T}_l^\dagger \gamma_5 D \mathcal{T}_l
\end{align}
which is just equivalent to a Galerkin coarsening of $Q$ instead, where $\mathcal{R}_l = \mathcal{T}_l^\dagger$. Thus coarsening $Q$ is a natural choice, given that $\phi_i$ are low modes of $Q$.
However, naively using the Petrov-Galerkin approach, \cref{eq:PG}, results in a coarse operator that may have arbitrarily small in magnitude eigenvalues, as discussed in \cref{appendix:spectra}.
One solution is to preserve chiral degrees of freedom on the coarse grids by
explicitly preserving chiral indices from the fine grid as follows.
This property makes it irrelevant whether we coarsen $D$ or $Q$, because then the coarse Dirac operator on level $l$ is $\gamma_5$-Hermitian with respect to the coarse $\gamma_5$-operator defined as $(\gamma_5)_l = \mathcal{R}_l \gamma_5 \mathcal{T}_l$. An additional effect of this is that the projector $P_l = \mathcal{T}_l \mathcal{R}_l$ maintains chirality of a state when projecting,
\begin{equation}
[P_l, \gamma_5] = 0 \implies \left( (\gamma_5)_l D_l \right)^\dagger = (\gamma_5)_l D_l,
\end{equation}
and we have
\begin{align}
\mathcal{T}_l Q^{-1}_l \mathcal{R}_l \gamma_5 = \mathcal{T}_l D^{-1}_l \mathcal{R}_l
\qquad
\text{and}
\qquad
Q_l = (\gamma_5 D)_l = (\gamma_5)_l D_l.
\end{align}
This can be achieved, for instance, by extending the set of $\Nc$ low modes of $Q$ before block-projection as follows
\begin{align}
\{\phi_i\}_{i=0}^{N_c-1} &\longmapsto \{\phi_i\}_{i=0}^{N_c-1} \cup \{\gamma_5 \phi_i\}_{i=0}^{N_c-1}
\label{eq:chirality:A} 
\end{align}
or by the equivalent version we choose in our prescription
\begin{align}
\{\phi_i\}_{i=0}^{N_c-1} &\longmapsto \{P_0 \phi_i\}_{i=0}^{N_c-1} \cup \{P_1 \phi_i\}_{i=0}^{N_c-1}
\label{eq:chirality:B}
\end{align}
where $P_\alpha = \tfrac{1}{2}(1+(-1)^\alpha\gamma_5)$ are the chiral projectors. Here, we assume that $\gamma_5 \phi_i$ is not proportional to any $\phi_j$ for all $i,j$, a requirement that will usually be satisfied in practice. Then the resulting subspace has dimension $2N_c$. After orthonormalizing the sets of low modes defined in the r.h.s.~of \cref{eq:chirality:A,eq:chirality:B}, the two extended sets are equivalent up to a basis transformation.
The set \cref{eq:chirality:A} has a nice interpretation, because is is the set of left and right singular vectors of $D$ as motivated above.
On the other hand, the coarse $\gamma_5$ matrix keeps the same form as the fine one, when choosing the set \cref{eq:chirality:B}.
We have used the extended set defined in \cref{eq:chirality:B} for all LMA and MG-LMA estimators in this work.

This leads to a doubling of the dimension of the coarse-grid subspace. 
The increase in subspace dimension is compensated by a positive impact of the imposed chirality preservation to the condition numbers of the coarse grid operators, see \cref{appendix:spectra}. When coarsening the Dirac operator $D$ directly, instead of $Q$, we also observe an impact on the variance contribution (see \cref{fig:chirality}).
Alternatively one could preserve all four spins, $[\projector, \gamma_{\mu}] = 0$, by
\begin{align}
\{\phi_i\}_{i=1}^{N_c}
&\longmapsto
     \{\gamma_0 \phi_i\}_{i=1}^{N_c}
\cup \{\gamma_1 \phi_i\}_{i=1}^{N_c}
\cup \{\gamma_2 \phi_i\}_{i=1}^{N_c}
\cup \{\gamma_3 \phi_i\}_{i=1}^{N_c},
\end{align}
for instance, but we have not seen any further improvements by doing this. The variance contribution and the condition number remains the same, but the subspace dimension is further increased by a factor of two.

\begin{figure}[!t]
    \centering
    \includegraphics[width=1.0\linewidth]{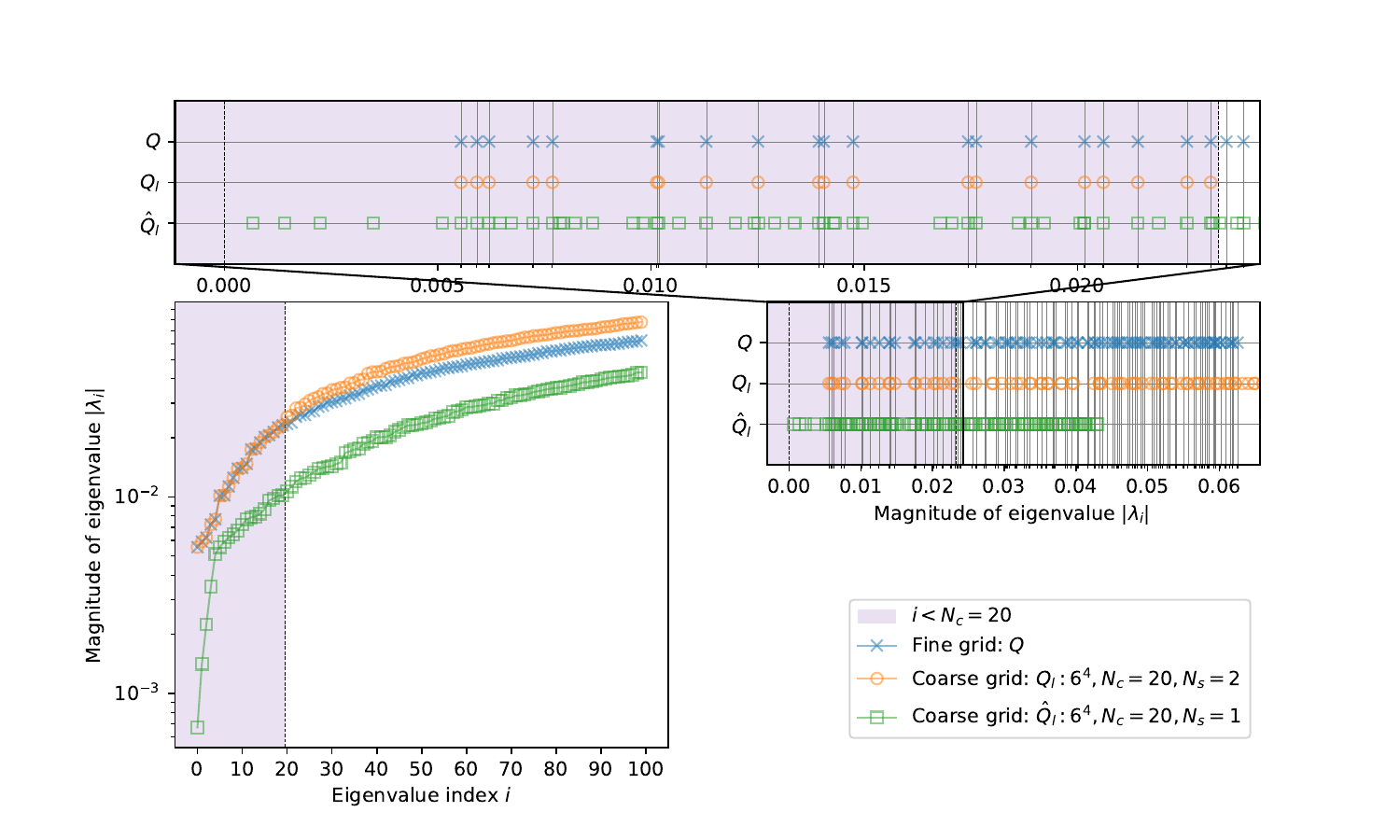}
    \caption{Lowest $100$ eigenvalues (in magnitude) of the Hermitian Dirac
    operator $Q$ and its coarse grid variants (lower left). In the lower right
panel, the eigenvalues of the operators are plotted on the x-axis with a grey
vertical line at every fine grid eigenvalue (blue crosses). The upper panel is a
zoomed version of the lower right panel in the relevant region of $i < N_c=20$.
We observe perfect alignment of the lowest $\Nc$ eigenvalues if chirality is
preserved.}
    \label{fig:spectra}
\end{figure}

\section{Study of the spectra of coarse and fine operators}
\label{appendix:spectra}

We observe further important implications of chirality preservation \cite{babich2010}, namely the eigenvalue spectra of coarse grid Dirac operators exhibit lower modes than their fine grid ones if chirality is not preserved, see \cref{fig:spectra}.
We plotted the $100$ lowest in magnitude eigenvalues of the fine- and coarse-grid Hermitian Dirac operators with different coarsening strategies.
The coarse operator (circle symbols) preserves chirality as defined in \cref{eq:chirality:B}, whereas the square symbols represent a coarse operator, where the coarse spin degree of freedom is trivial.
On the right and top panel, we clearly see a perfect overlap of the $\Nc=20$ lowest modes in the chiral-preserving case (circles) and some spurious low eigenvalues in the other case (squares).
This makes coarse operators without chirality increasingly ill-conditioned and hard to invert.
Very quickly, we observed cases where coarse grid Dirac operators showed
larger condition numbers than fine grid ones, and thus violate conclusions
drawn in \cref{appendix:cond,appendix:mrhs}, unless we preserve chirality on the coarse subspaces.

\begin{figure}[!t]
    \centering
    \includegraphics[width=0.6\linewidth]{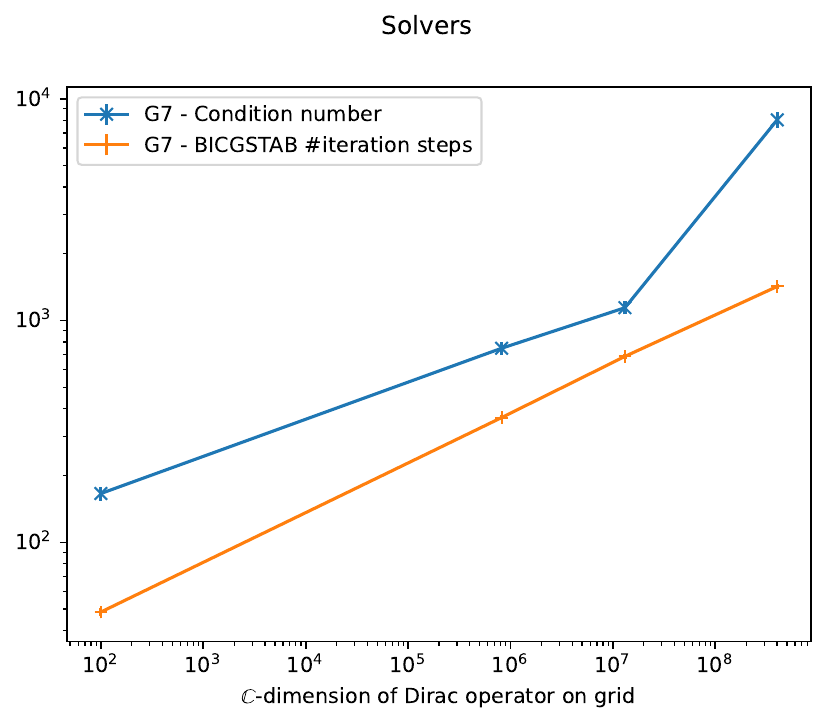}
    \caption{Condition number (blue crosses) and iteration count (yellow plusses) of Dirac operators on different grids whose dimension over $\mathbb{C}$ is plotted on the x-axis. All coarse operators considered have explicit chiral degrees of freedom.}
    \label{fig:coarse:cond}
\end{figure}

\section{Condition numbers of coarse-grid operators}

\label{appendix:cond}

We have investigated the condition number of the coarse-grid operators numerically in \cref{fig:coarse:cond} by defining the four operators on one of the lattices (G7).
The operators were generated using $N_c=50$ low modes, chirality preservation for $N_s=2$ spin d.o.f. (see \cref{appendix:chirality}) and different coarse grids of sizes of $1^4$, $16 \times 8^3$, $32 \times 16^3$, $128 \times 64^3$ (left to right).
The leftmost data point corresponds to an operator in an LMA-scenario (no explicit blocking), the rightmost data point is the fine grid Dirac operator, intermediate points are non-trivial MG-levels.
The plot shows the dimensionality over $\mathbb{C}$ (x-axis) vs. the estimated condition number (blue crosses) as well as the iteration count of an unpreconditioned BiCGSTAB\footnote{Any other Krylov subspace solver would exhibit the same quantitative picture.} solve with a random right hand side (yellow plusses).
We observe that coarse grid Dirac operators are better conditioned and thus require fewer solver iterations than the operator defined on the fine grid.

\clearpage

\section{Additional measurements on \texorpdfstring{$L=2.1\mathrm{fm}$ and $3.1\mathrm{fm}$ lattices}{L=2.1fm and 3.1fm lattices}}

\label{appendix:plots}

Here, we present additional measurements on the lattices not discussed in the main text (see \cref{sec:results}). These are performed on E7 and F7 lattices with spatial extents $64  \times 32^3$ and $96  \times 48^3$ respectively. 

\subsection{E7}
\begin{figure}[htbp]
    \centering
    \includegraphics[width=1.0\linewidth]{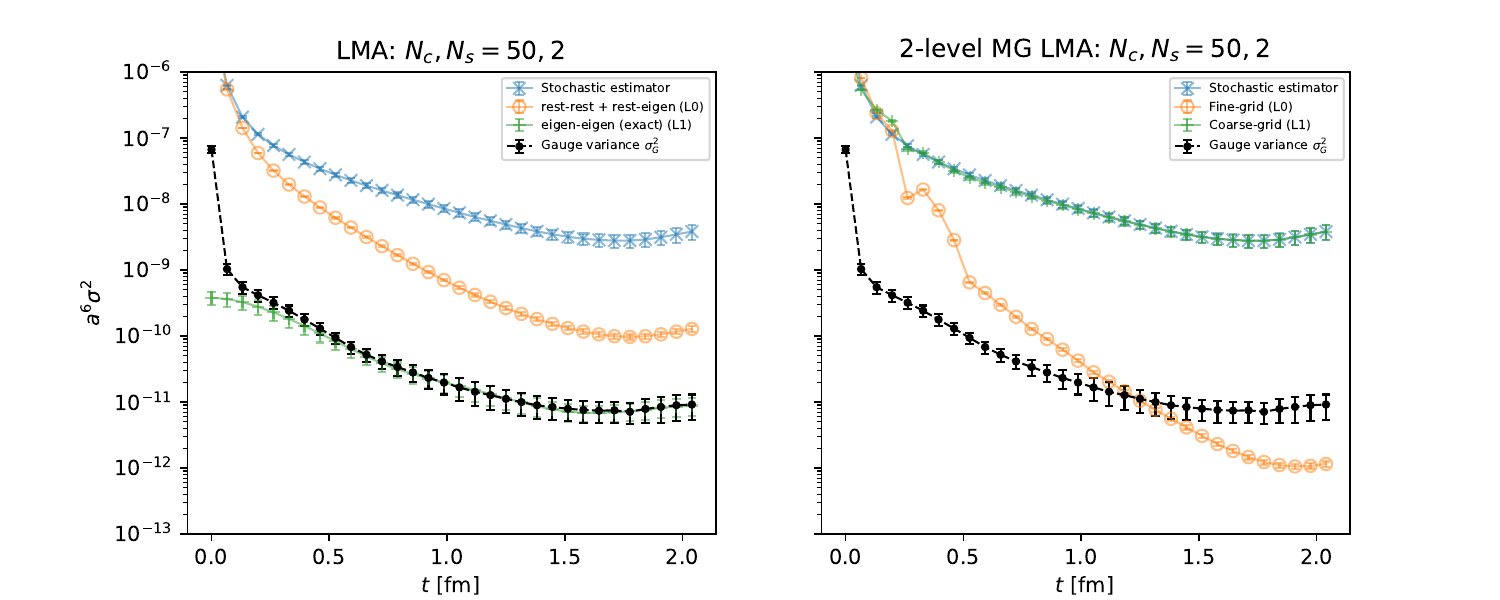}
    \includegraphics[width=1.0\linewidth]{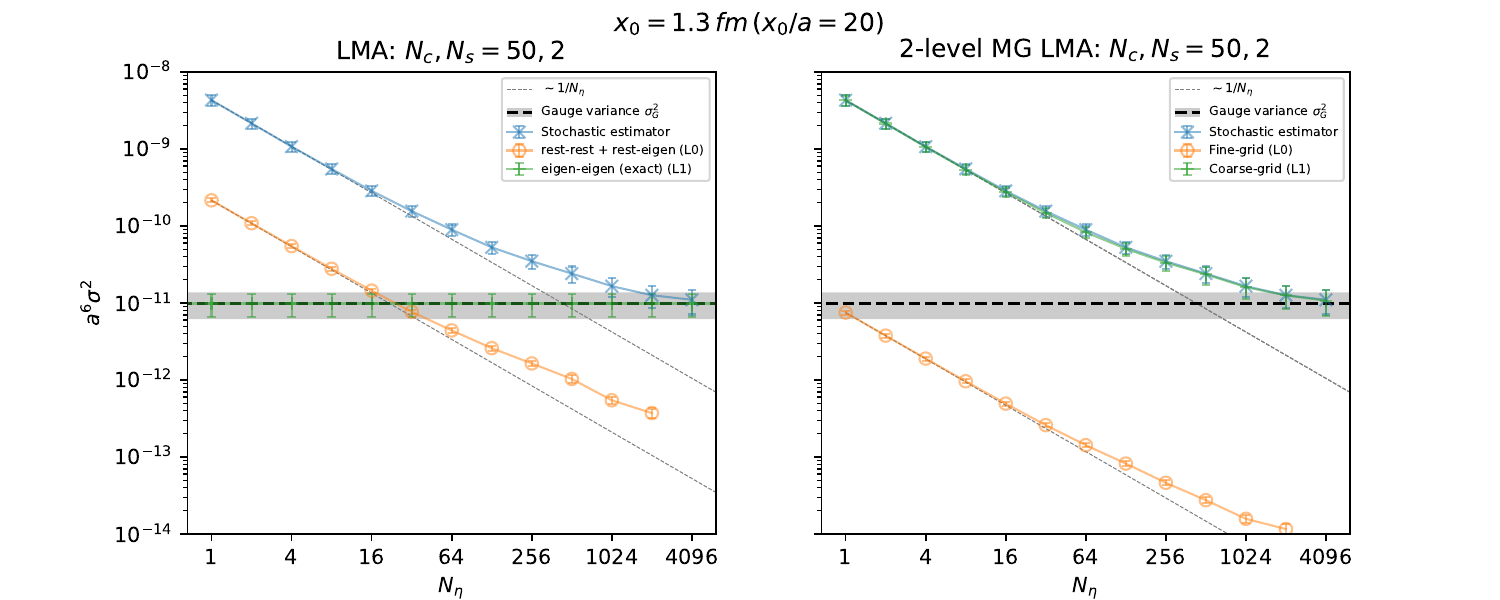}
    \caption{Top: Absolute variances of the lattice E7 for LMA (left) and MG LMA (right) to the vector correlator \cref{eq:cdecomp} with one stochastic source for each term. The black line is the gauge variance. The standard stochastic estimator (blue) appears in both plots for reference.
Bottom: Absolute variances of the lattice E7 for LMA (left) and MG LMA (right) against the number of stochastic sources $\Nsrc$ on time separation $t = 1.3\,\mathrm{fm}$. The black line is the gauge variance. The standard stochastic estimator (blue) appears in both plots for reference.}
    \label{fig:E7:var}
\end{figure}
\clearpage

\subsection{F7}
\begin{figure}[htbp]
    \centering
    \includegraphics[width=1.0\linewidth]{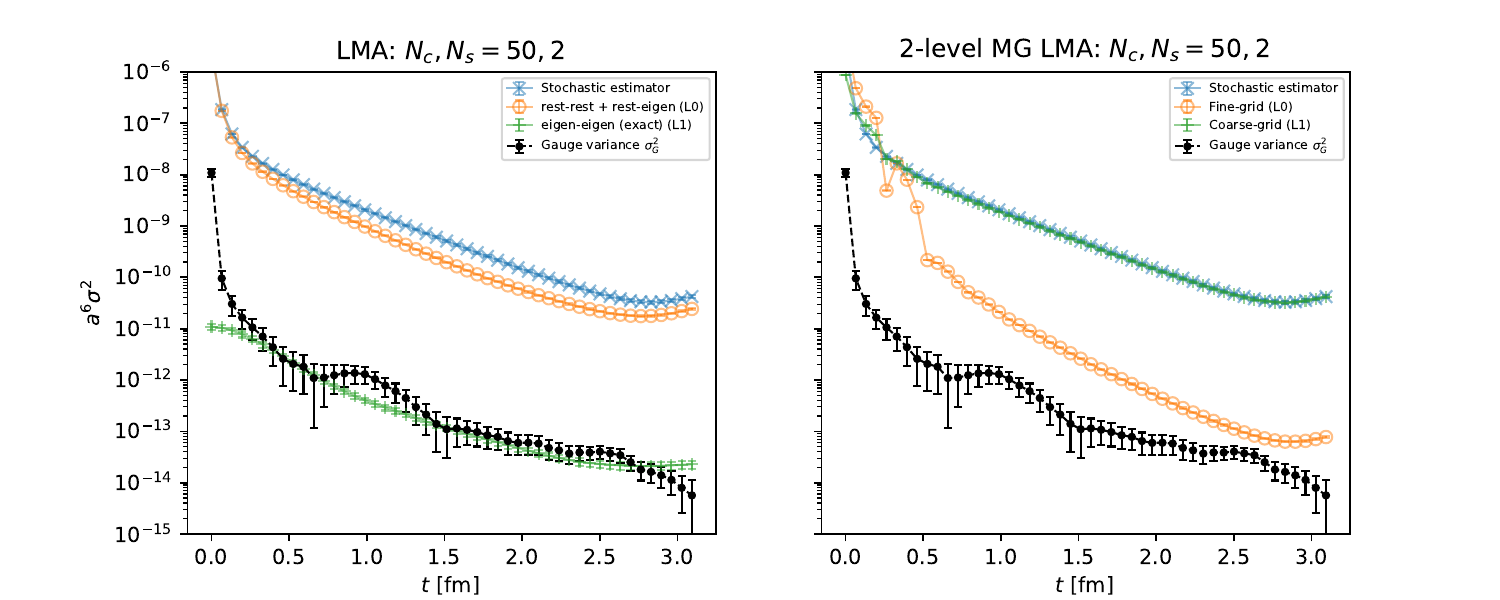}
    \includegraphics[width=1.0\linewidth]{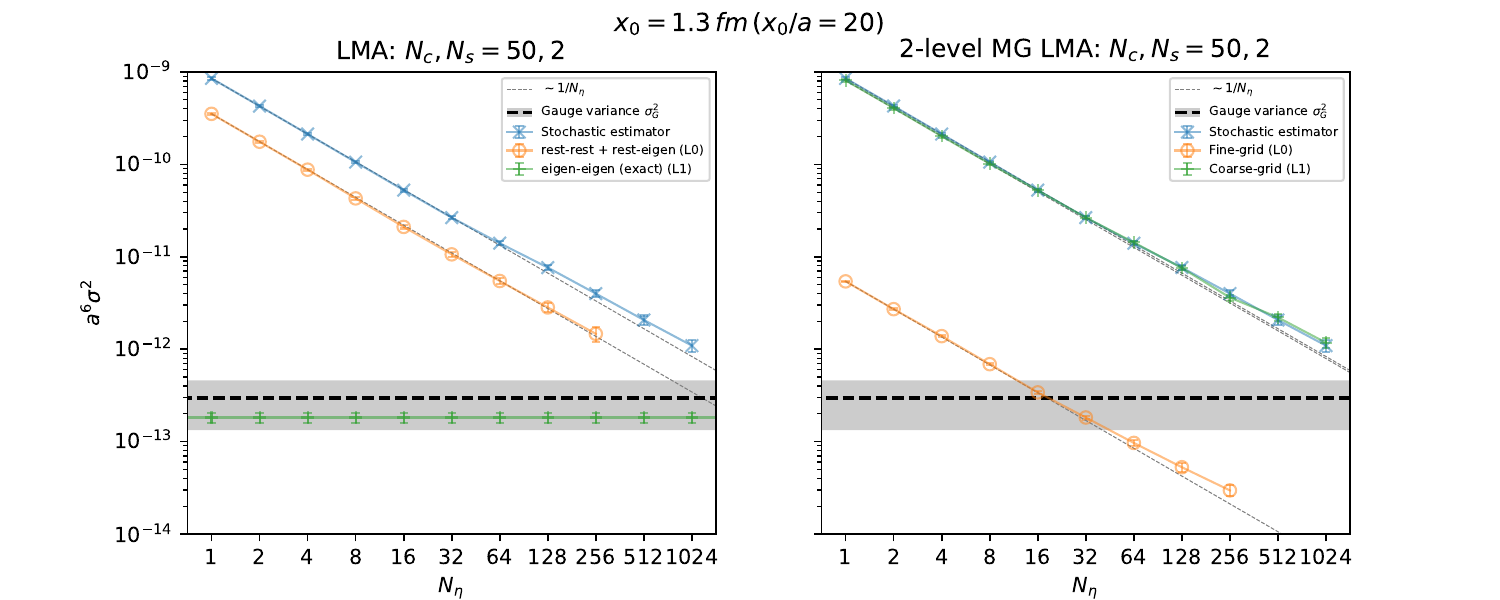}
    \caption{Top: Absolute variances of the lattice F7 for LMA (left) and MG LMA (right) to the vector correlator \cref{eq:cdecomp} with one stochastic source for each term. The black line is the gauge variance. The standard stochastic estimator (blue) appears in both plots for reference.
Bottom: Absolute variances of the lattice F7 for LMA (left) and MG LMA (right) against the number of stochastic sources $\Nsrc$ on time separation $t = 1.3\,\mathrm{fm}$. The black line is the gauge variance. The standard stochastic estimator (blue) appears in both plots for reference.}
    \label{fig:F7:var}
\end{figure}
\clearpage

\printbibliography
\end{document}